\documentstyle[11pt,epsf]{article}

\makeatletter
\if@twoside
\else \oddsidemargin 00pt \evensidemargin 00pt
\fi
\def\section{\@startsection {section}{1}{\z@}{+6.0ex plus +1ex minus
 +.2ex}{2.8ex plus .2ex}{\large\bf}}
\def\subsection{\@startsection {subsection}{2}{\z@}{+3.0ex plus +1ex
minus +.2ex}{2.3ex plus .2ex}{\normalsize\bf}}
\def\subsubsection{\@startsection{subsubsection}{3}{\z@}{+2.5ex plus
+1ex minus +.2ex}{1.5ex plus .2ex}{\normalsize\bf}}
 
\def\theequation{\thesection.\arabic{equation}}
\@addtoreset{equation}{section}

\@addtoreset{figure}{section}

\@addtoreset{table}{section}
 
\def\appendix{\par
 \setcounter{section}{0} \setcounter{subsection}{0}
 \def\thesection{\Alph{section}}}
 
\expandafter\ifx\csname mathrm\endcsname\relax\def\mathrm#1{{\rm #1}}\fi
\@ifundefined{mathrm}{\def\mathrm#1{{\rm #1}}}{\relax}
\makeatother
 
\def\beq{\begin{equation}}
\def\eeq{\end{equation}}
\def\beqar{\begin{eqnarray}}
\def\eeqar{\end{eqnarray}}
\def\bma{\begin{displaymath}}
\def\ema{\end{displaymath}}
\def\barr#1{\begin{array}{#1}}
\def\earr{\end{array}}
\def\bit{\begin{itemize}}
\def\eit{\end{itemize}}
\def\bfi{\begin{figure}}
\def\efi{\end{figure}}
\def\btab{\begin{table}}
\def\etab{\end{table}}
\def\bce{\begin{center}}
\def\ece{\end{center}}

\makeatletter
\def\refe#1{\let\@refa\@empty
           \@for\@ref:=#1\do{\@refa\def\@refa{,\penalty\@m\ }%
                             \ref{\@ref}}}
\makeatother

\def\mathswitchr#1{\relax\ifmmode{\mathrm{#1}}\else$\mathrm{#1}$\fi}

\def\mathswitch#1{\relax\ifmmode#1\else$#1$\fi}

\makeatletter
\let\@eqnsel = \hfil
 
\def\eqnarray{\stepcounter{equation}\let\@currentlabel=\theequation
\global\@eqnswtrue
\global\@eqcnt\z@\tabskip\@centering\let\\=\@eqncr
$$\halign to \displaywidth\bgroup\hskip\@centering
  $\displaystyle\tabskip\z@{##}$\@eqnsel&\global\@eqcnt\@ne
  \hskip 2\arraycolsep \hfil${##}$\hfil
  &\global\@eqcnt\tw@ \hskip 2\arraycolsep $\displaystyle\tabskip\z@{##}$\hfil
   \tabskip\@centering&\llap{##}\tabskip\z@\cr}
\makeatother

\newcommand{\sw}{s_w}
\newcommand{\cw}{c_w}
\newcommand{\twz}{\frac{\sw^2}{\cw^2}}

\newcommand{\Mh}{M_{h^0}}
\newcommand{\mh}{m_{h^0}}
\newcommand{\MH}{M_{H^0}}
\newcommand{\mH}{m_{H^0}}
\newcommand{\MP}{M_{H^+}}
\newcommand{\MA}{M_{A^0}}
\newcommand{\MS}{M_{SUSY}}

\newcommand{\mt}{m_{top}}
\newcommand{\mtr}{m_{\tilde{t}_R}}
\newcommand{\mtl}{m_{\tilde{t}_L}}
\newcommand{\mts}{m_{\tilde{t}}}
\newcommand{\mf}{m_f}
\newcommand{\mfr}{m_{\tilde{f}_R}}
\newcommand{\mfl}{m_{\tilde{f}_L}}

\newcommand{\mfsf}{m_{f'}}
\newcommand{\mfsr}{m_{\tilde{f'}_R}}
\newcommand{\mfsl}{m_{\tilde{f'}_L}}

\newcommand{\mcpi}{m_{\tilde{\chi}_i^+}}
\newcommand{\mcpj}{m_{\tilde{\chi}_j^+}}
\newcommand{\mcpk}{m_{\tilde{\chi}_k^+}}
\newcommand{\mcni}{m_{\tilde{\chi}_i^0}}
\newcommand{\mcnj}{m_{\tilde{\chi}_j^0}}
\newcommand{\mcnk}{m_{\tilde{\chi}_k^0}}
\newcommand{\hAA}{$h^0A^0A^0$\hspace{2mm}}
\newcommand{\HAA}{$H^0A^0A^0$\hspace{2mm}}
\newcommand{\htAA}{$h^0 \to A^0A^0$\hspace{2mm}}

\newcommand{\BS}{\begin{samepage}}
\newcommand{\ES}{\end{samepage}}
\newcommand{\KL}{\left(}
\newcommand{\KR}{\right)}
\newcommand{\KKL}{\left[}
\newcommand{\KKR}{\right]}

\newcommand{\BL}{\lbrack}
\newcommand{\BR}{\rbrack}
\newcommand{\VL}{\left( \begin{array}{c}}
\newcommand{\VR}{\end{array} \right)}
\newcommand{\ML}{\left( \begin{array}{cc}}
\newcommand{\MLv}{\left( \begin{array}{cccc}}
\newcommand{\MR}{\end{array} \right)}

\newcommand{\hc}{\mbox {h.c.}}
\newcommand{\re}{\mbox {Re}}
\newcommand{\OP}{\omega_+}
\newcommand{\OM}{\omega_-}
\newcommand{\omt}{\omega_t}
\newcommand{\aeff}{\alpha_{eff}}
\newcommand{\Tb}{\tan \beta\hspace{1mm}}
\newcommand{\TQb}{\tan^2 \beta\hspace{1mm}}
\newcommand{\CTb}{\cot \beta\hspace{1mm}}
\newcommand{\CTQb}{\cot^2 \beta\hspace{1mm}}
\newcommand{\Sb}{\sin \beta\hspace{1mm}}
\newcommand{\SQb}{\sin^2\beta\hspace{1mm}}
\newcommand{\Cb}{\cos \beta\hspace{1mm}}

\newcommand{\Sa}{\sin \alpha\hspace{1mm}}
\newcommand{\SQa}{\sin^2\alpha\hspace{1mm}}
\newcommand{\Ca}{\cos \alpha\hspace{1mm}}
\newcommand{\CQa}{\cos^2\alpha\hspace{1mm}}
\newcommand{\Sab}{\sin (\alpha + \beta)\hspace{1mm}}
\newcommand{\Cab}{\cos (\alpha + \beta)\hspace{1mm}}
\newcommand{\Sba}{\sin (\beta - \alpha)\hspace{1mm}}
\newcommand{\Cba}{\cos (\beta - \alpha)\hspace{1mm}}
\newcommand{\SZa}{\sin 2\alpha\hspace{1mm}}

\newcommand{\CZa}{\cos 2\alpha\hspace{1mm}}
\newcommand{\CQZa}{\cos^2 2\alpha\hspace{1mm}}
\newcommand{\SZb}{\sin 2\beta\hspace{1mm}}
\newcommand{\SQZb}{\sin^2 2\beta\hspace{1mm}}
\newcommand{\CZb}{\cos 2\beta\hspace{1mm}}
\newcommand{\CQZb}{\cos^2 2\beta\hspace{1mm}}
\newcommand{\CDZb}{\cos^3 2\beta\hspace{1mm}}
\newcommand{\CQba}{\cos^2 (\beta - \alpha)\hspace{1mm}}
\newcommand{\SQba}{\sin^2 (\beta - \alpha)\hspace{1mm}}
\newcommand{\SDba}{\sin^3 (\beta - \alpha)\hspace{1mm}}
\newcommand{\CQab}{\cos^2 (\alpha + \beta)\hspace{1mm}}

\newcommand{\SDab}{\sin^3 (\alpha + \beta)\hspace{1mm}}
\newcommand{\vorn}{\hspace{-0.6cm}}
\newcommand{\BE}{\begin{equation}}
\newcommand{\EE}{\end{equation}}
\newcommand{\BEA}{\begin{eqnarray}}
\newcommand{\EEA}{\end{eqnarray}}
\newcommand{\BSUB}{\begin{subeqnarray}}
\newcommand{\ESUB}{\end{subeqnarray}}
\newcommand{\dline}[8]{\multiput(#1,#2)(#3,#4){#5}{\line(#6,#7){#8}}}
\newcommand{\GeV}{\hspace{1mm}\mbox{GeV}}

\begin{document}

\thispagestyle{empty}
\hfill KA-TP-17-1995

\hfill hep-ph/9602318

\vspace{2cm}
\begin{center}

{\Large \bf
The decay ${\mbox h}^0 \to {\mbox A}^0{\mbox A}^0$: \\ 
\vspace{5mm} 
a complete 1-loop calculation in the MSSM}

\vspace{2cm}

{\large \sc S. Heinemeyer and W. Hollik}

\vspace{2cm}

{\sl Institut f\"ur Theoretische Physik}

{\sl Universit\"at Karlsruhe}

{\sl Kaiserstra\ss{}e 12}

{\sl 76128 Karlsruhe, Germany}

\end{center}

\vspace{1cm}

\begin{abstract}
\noindent
In the minimal supersymmetric standard model
 the decay \htAA of the light neutral scalar $h^0$
 is kinematically
allowed for low values of $\tan\beta$ when radiative corrections
to the neutral Higgs masses are taken into account.
The width of this decay mode is revisited on the basis of a complete
1-loop diagrammatic calculation. We give the analytical expressions
and numerical results and compare them with the corresponding ones 
from the simpler and compact approximations of the effective
potential method and the renormalization group approach. 
\end{abstract}

\newpage

\setcounter{page}{1}

\section{Introduction}
\noindent
The search for Higgs bosons is a basic challenge for particle physics
with the aim to shed light on the mechanism of electroweak symmetry
breaking.
The Minimal Supersymmetric Standard Model (MSSM) predicts at least
one light neutral scalar Higgs particle. In order to
experimentally  detect possible signals of Higgs bosons and to 
trace back as far as possible the physical origin of a produced
scalar particle, detailed studies of production and decay processes
of Higgs bosons are required.

The Higgs sector of the MSSM consists of two neutral CP-even
states $h^0, H^0$, one neutral CP-odd state $A^0$, and a pair of
charged bosons $H^\pm$. 
At tree level, there are only two independent input parameters and thus 
the Higgs masses  are strongly correlated.
As a special consequence,
the bound $M_{h^0} < \MA$ kinematically forbids the decay mode
\htAA of the lightest neutral scalar boson $h^0$.
As has been found few years ago and worked out in various stages 
of approximations 
\cite{habhem,elrizwi,brignole,barbieri,brignole2,yamada,haa,cpr,ds,ds1},
radiative corrections in the MSSM
Higgs sector are large, proportional to $\mt^4$ at 1-loop
 in their leading terms.
They  significantly modify the tree level
relations between masses and couplings and hence have to be taken into
account for phenomenological studies and for interpretation of the results
from experimental searches.

A specific consequence of the large radiative corrections to the $h^0$ mass
is the possibility of having $M_{h^0} > 2 M_{A^0}$, which makes the
decay \htAA kinematically allowed.
In refs.~\cite{brignole,haa} it was shown that it can be the dominant 
decay mode of the $h^0$ and thus is of crucial importance for the
experimental search for supersymmetric Higgs bosons. The calculation of the
decay width  was performed in various approximations: in the
effective potential approach \cite{brignole} and by the use of
renormalization group equations \cite{haa}.
The expressions obtained in this way are
of impressive simplicity and thus most suitable for fast handling in
Monte Carlo studies. 

\smallskip
Besides these approximations, a diagrammatic calculation complete at the
1-loop level is desirable. It allows for the virtual contributions from all
particles from the MSSM spectrum and the momentum dependence of the
2- and 3-point functions. This method is the technically most complicated
one, but also most accurate at the 1-loop level.
It provides a reference frame for checking the quality of
the compact approximate formulae and allows to study the full 
parameter dependence of the decay rate. In the analysis of 
experimental data for the MSSM Higgs search done
so far \cite{aleph} only
 the available approximations were used.

\smallskip
In this article the width for the decay mode
\htAA  of the light MSSM neutral scalar boson
is revisited on the basis of a complete 1-loop diagrammatic
calculation. In addition, we compare the results with the 
corresponding ones from the  simpler approximations and discuss 
the typical size of the differences.
The calculation is performed in the on-shell renormalization scheme in the
version worked out in ref.~\cite{ds}.
 Section 2 contains a brief summary of the structure of the propagator
corrections and corrections to the neutral Higgs masses.
The set of vertex corrections to the \hAA vertex is given in
section 3 with the explicit analytical expressions in the appendix.
Results for the decay widths and branching ratios are presented and
discussed in sections 4 and 5. The appendices provide
all conventions and necessary formulae.   

\pagebreak

\section{The MSSM Higgs sector}
\subsection{Tree-level structure}
\noindent
The  Higgs potential of the MSSM reads \cite{hhg}
\BEA
\label{Higgspot}
V =& &m_1^2 H_1\bar{H}_1 + m_2^2 H_2\bar{H}_2 + m_{12}^2 (\epsilon_{ab}
      H_1^aH_2^b + \hc) \\ \nonumber
   &+&\frac{1}{8}(g'^2 + g^2)(H_1\bar{H}_1 - H_2\bar{H}_2)^2
      -\frac{g^2}{2}|H_1\bar{H}_2|^2,
\EEA
where $m_i, m_{12}$ are mass parameters, and
$g, g'$  the $SU(2)$ and $U(1)$ gauge couplings.

\noindent
$H_1$ and $H_2$ are decomposed  in the following way:
\BEA
H_1 &=& \VL H_1^1 \\ H_1^2 \VR = \VL v_1 + (\phi_1^{*0} + i\chi_1^{*0})
                                 /\sqrt2 \\ \phi_1^- \VR ,\\ \nonumber
H_2 &=& \VL H_2^1 \\ H_2^2 \VR = \VL \phi_2^+ \\ v_2 + (\phi_2^0 
                                     + i\chi_2^0)/\sqrt2 \VR.
\EEA
The vacuum expectation values $v_1$ and $v_2$ define
the angle $\beta$ in terms of the ratio
\BE
\Tb = \frac{v_2}{v_1}.
\EE
In order to obtain the CP-even neutral
 mass eigenstates,  the rotation 
\BEA
\VL H^0 \\ h^0 \VR &=& \ML \Ca & \Sa \\ -\Sa & \Ca \MR 
\VL \phi_1^0 \\ \phi_2^0 \VR ,\\ \nonumber
%\VL G^0 \\ A^0 \VR &=& \ML \Cb & \Sb  \\ -\Sb  & \Cb \MR 
%\VL \chi_1^0 \\ \chi_2^0 \VR ,\\ \nonumber
%\VL G^+ \\ H^+ \VR &=& \ML \Cb & \Sb  \\ -\Sb  & \Cb \MR 
%\VL \phi_1^+ \\ \phi_2^+ \VR. 
\EEA
is performed. 
The spectrum consists of: 
\BEA
\mbox{2 neutral bosons, CP=1} &:& h^0,H^0 \nonumber \\
\mbox{1 neutral boson, CP=$-$1} &:& A^0 \nonumber \\
\mbox{2 charged bosons} &:& H^+,H^- \nonumber \\
\mbox{3 unphysical Goldstone bosons} &:& G^0,G^+,G^- \nonumber
\EEA
The potential (\ref{Higgspot}) can be fixed with the help of  two  
independent parameters: $\Tb$ and
\BE
M_{A^0}^2 = -m_{12}^2(\Tb+\CTb),
\EE
where $M_{A^0}$ is
the mass of the $A^0$ boson.
The mixing angle of the $(h^0,H^0)$ system at tree level can be obtained from
\BE
\label{alphaborn}
\tan 2\alpha = \tan 2\beta \frac{\MA^2 + M_Z^2}{\MA^2 - M_Z^2},\quad
-\frac{\pi}{2} < \alpha \le 0,
\EE
and the masses of all other physical Higgs bosons are given  by
\BEA
\label{bornHiggsmassen}
m_{H^0,h^0}^2 &=& \frac{1}{2} \KKL M_A^2 + M_Z^2 \pm \sqrt{(M_A^2+M_Z^2)^2
                  -4M_Z^2M_A^2\cos^22\beta}\hspace{1mm}\KKR, \\ \nonumber
m_{H^\pm}^2 &=& M_A^2 + M_W^2.
\EEA
%
%At tree level the masses are constrained by
%\BEA
%\label{bornHiggsmassen2}
%m_{h^0} < \MA,M_Z < m_{H^0}, \qquad \MA, M_W < m_{H^\pm},\\ \nonumber
%m_{h^0}^2 + m_{H^0}^2 = \MA^2 + M_Z^2.
%\EEA
%
\subsection{Radiative corrections to the Higgs masses}
\noindent
The simple tree-level results are changed significantly
 when radiative corrections are 
taken into account
\cite{habhem,elrizwi,brignole,barbieri,brignole2,yamada,haa,cpr,ds}.
In this paper we make use of  the on-shell
 renormalisation scheme worked out in \cite{ds}.
The inverse of the $(h^0,H^0)$ propagator matrix
\BE
\Delta^{-1} = -i \ML k^2 - m_{h^0} + \hat{\Sigma}_{h^0}(k^2) & 
                     \hat{\Sigma}_{h^0H^0}(k^2) \\
                     \hat{\Sigma}_{h^0H^0}(k^2) & 
                     k^2 - m_{H^0} + \hat{\Sigma}_{H^0}(k^2) \MR ,
\EE
contains the renormalized self-energies
$\hat{\Sigma}_{h^0},\hat{\Sigma}_{H^0},\hat{\Sigma}_{h^0H^0}$,
documented in~\cite{ds}. 
The individual propagators in the matrix
\BE
\label{propmatrix}
\Delta = i \ML \Delta_{h^0} & \Delta_{h^0H^0} \\ \Delta_{h^0H^0}  &
               \Delta_{H^0} \MR
\EE
are given by
\BEA
\Delta_{h^0} &=& \frac{1}{k^2 - m_{h^0}^2 + \hat{\Sigma}_{h^0}(k^2) -
                 \frac{\hat{\Sigma}_{h^0H^0}^2(k^2)}{k^2 - m_{H^0}^2 +
                 \hat{\Sigma}_{H^0}(k^2)}},\\ \nonumber
\Delta_{H^0} &=& \frac{1}{k^2 - m_{H^0}^2 + \hat{\Sigma}_{H^0}(k^2) -
                 \frac{\hat{\Sigma}_{h^0H^0}^2(k^2)}{k^2 - m_{h^0}^2 +
                 \hat{\Sigma}_{h^0}(k^2)}},\\ \nonumber
\Delta_{h^0H^0} &=& \frac{-\hat{\Sigma}_{h^0H^0}(k^2)}
                 {(k^2 - m_{h^0}^2 + \hat{\Sigma}_{h^0}(k^2))
                  (k^2 - m_{H^0}^2 + \hat{\Sigma}_{H^0}(k^2)) 
                  - \hat{\Sigma}_{h^0H^0}^2(k^2)}.
\EEA
In these expressions,
 $\mh$ and $m_{H^0}$ are the formal parameters of 
(\ref{bornHiggsmassen}). 
The physical one-loop masses are determined by
the pole positions of the matrix (\ref{propmatrix}). They can be
obtained as the solutions of the equation
\BE
(k^2 -m_{h^0}^2 + \hat{\Sigma}_{h^0}(k^2))(k^2 - m_{H^0}^2 + \hat{\Sigma}_
{H^0}(k^2)) - (\hat{\Sigma}_{h^0H^0}(k^2))^2 = 0.
\EE
The physical Higgs masses are denoted by $\Mh$ and $\MH$ in order to 
distinguish them from the formal tree-level masses (\ref{bornHiggsmassen}).

\begin{figure}[htbp]
\begin{picture}(150,90)
\put(20,5){\epsffile[100 530 577 775]{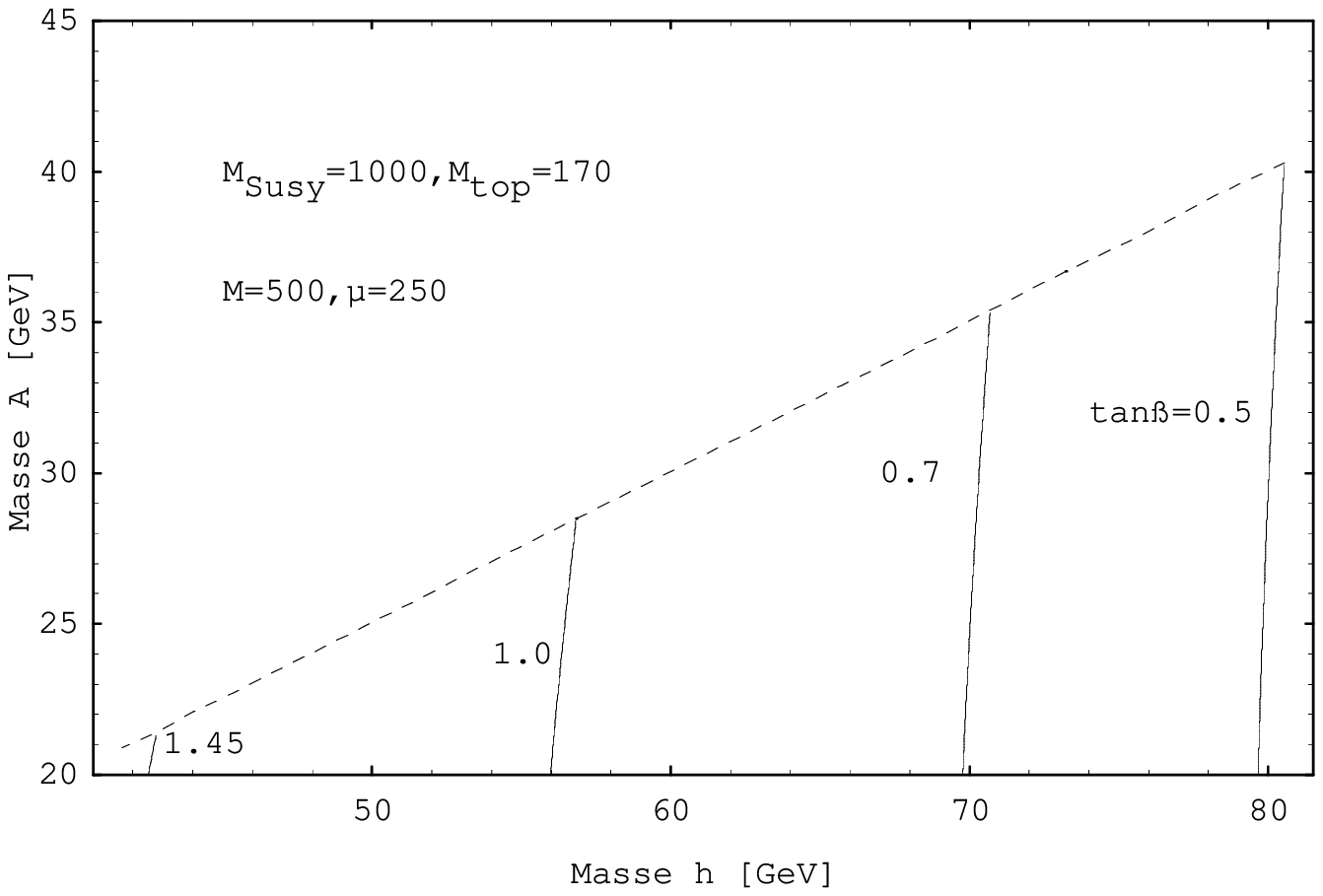}}
\put(80,2){\makebox(0,0)[t]{$\Mh [\GeV]$}}
\put(8,50){\makebox(0,0)[l]{$\MA$}}
\put(8,45){\makebox(0,0)[l]{$[\GeV]$}}
\end{picture}
\caption{$\Mh - \MA$-dependence where $h^0 \to A^0A^0$
         is kinematicaly allowed}
\label{mhmatb}
\end{figure}

\smallskip
As a consequence of the sizeable loop contributions
 there is a part of the parameter space with $\Mh > 2\MA$, where thus
the decay \htAA is kinematically allowed.
The correlation of $\Mh, \MA$ and $\Tb$ 
for allowed  \htAA decays
is shown in Figure \ref{mhmatb} (all masses in GeV, the same applies
for all figures) for a typical set of parameters.
 $\MS$ is the scalar mass parameter, as 
 explained in (\ref{msusy}).

\epsfclipon
\unitlength1mm
\epsfclipoff

\subsection{$\mt^4$-approximation}
\noindent
For the Higgs masses, the dominant 1-loop
 contributions are the leading terms
$\sim\mt^4$\cite{ds}. In the simplest approximation, 
assuming unmixed left and right stop squarks,
the renormalised  
self energies can be expressed as
\BEA
\hat{\Sigma}_{h^0}(k^2) &=& -\omt\CQa, \nonumber \\
\hat{\Sigma}_{H^0}(k^2) &=& -\omt\SQa, \nonumber \\
\hat{\Sigma}_{h^0H^0}(k^2) &=& -\omega_t\Sa\Ca, \nonumber \\
\EEA
with ($N_C$ is the number of the colors, $\sw^2=1-M_W^2/M_Z^2$)
\BE
\omega_t = \frac{2N_Ce^2\mt^4}{(4\pi)^2\sw^2M_W^2\SQb}
\log(\frac{m_{\tilde{t}_L} m_{\tilde{t}_R}}{\mt^2}).
\EE
Accordingly,
equation (\ref{bornHiggsmassen}) is modified in the following way:
\BE
\label{omegatHiggsmassen}
M_{H^0,h^0}^2=\frac{\MA^2+M_Z^2+\omt}{2}\pm\sqrt{\frac{(\MA^2+M_Z^2)^2
               +\omt^2}{4}-\MA^2M_Z^2\CQZb
               +\frac{\omt\CZb}{2}(\MA^2-M_Z^2)}.
\EE

\section{Vertex corrections for ${\mbox h}^0 \to {\mbox A}^0 {\mbox A}^0$}
\subsection{Complete 1-loop terms}
\noindent 
The elementary
 \hAA -vertex arises from the following  term
 of the Higgs potential 
(\ref{Higgspot}):
\BE
\label{haapot}
\frac{1}{8}(g'^2 + g^2)(H_1\bar{H}_1 - H_2\bar{H}_2)^2
\EE
which leads to
\BE
\label{bornvertex}
\frac{-igM_Z}{2\cw}\CZb\Sab \quad =:iT_0, \;\;\;\;
 c_w = \cos\theta_W .
\EE
The \HAA -vertex is obtained by the 
substitution $\Sab \to -\Cab$.

\bigskip 
%\subsection{Radiative corrections}
At the one-loop level the \hAA -vertex is given by 
$T_0 + \Delta T_{h^0}$
 with $T_0$ from (\ref{bornvertex}) and the renormalized
vertex correction $\Delta T_{h^0}$.
 This expression consists of the sum
of all vertex correction diagrams and the counter term:
\BE
\label{deltatkh}
i\Delta T_{h^0} = \sum_{i} \Delta T_{h^0,i} + CT_{h^0}
\EE
The \HAA -vertex  at the one loop
level is obtained in an analogous way:
\BE
\label{deltatgh}
i\Delta T_{H^0} = \sum_{i} \Delta T_{H^0,i} + CT_{H^0}
\EE
All diagrams contributing to the vertex corrections are shown in 
Figures~\ref{vertexdiag1} and \ref{vertexdiag2}. Their analytical expressions 
 are explicitly given in appendix B
(for \hAA ).

\epsfclipon
\unitlength1mm
\begin{figure}[hbt]
\begin{picture}(150,120)
%\put(0,0){\framebox(150,120)[t]{}}
\put(000,085){\epsffile[188 415 302 486]{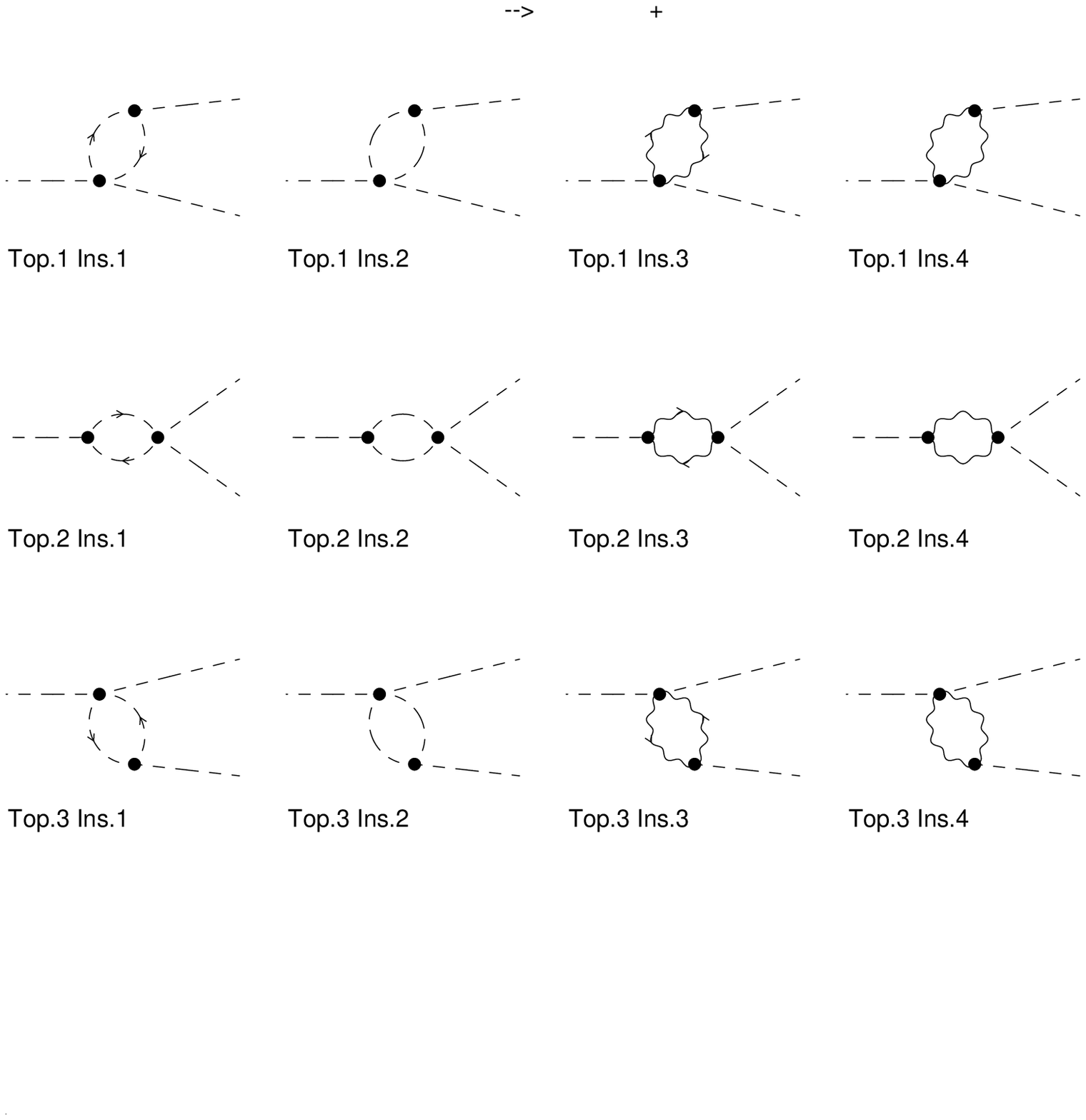}}
\put(019,102){\makebox(0,0)[b]{$h^0,H^0$}}
\put(019,092){\makebox(0,0)[t]{$h^0,H^0$}}
\put(000,099){\makebox(0,0)[l]{$h^0,H^0$}}
\put(039,089){\makebox(0,0)[l]{$A^0$}}
\put(039,106){\makebox(0,0)[l]{$A^0$}}
\put(025,080){\makebox(0,0)[b]{1.}}
\put(050,085){\epsffile[188 415 302 486]{top234.eps}}
\put(069,102){\makebox(0,0)[b]{$A^0,G^0$}}
\put(069,092){\makebox(0,0)[t]{$A^0,G^0$}}
\put(050,099){\makebox(0,0)[l]{$h^0,H^0$}}
\put(089,089){\makebox(0,0)[l]{$A^0$}}
\put(089,106){\makebox(0,0)[l]{$A^0$}}
\put(075,080){\makebox(0,0)[b]{2.}}
\put(100,085){\epsffile[072 415 185 486]{top234.eps}}
\put(119,102){\makebox(0,0)[b]{$H^+,G^+$}}
\put(119,092){\makebox(0,0)[t]{$H^-,G^-$}}
\put(100,099){\makebox(0,0)[l]{$h^0,H^0$}}
\put(139,089){\makebox(0,0)[l]{$A^0$}}
\put(139,106){\makebox(0,0)[l]{$A^0$}}
\put(125,080){\makebox(0,0)[b]{3.}}
\put(000,045){\epsffile[072 415 185 486]{top234.eps}}
\put(019,062){\makebox(0,0)[b]{$\tilde{f}_R$}}
\put(019,052){\makebox(0,0)[t]{$\tilde{f}_R$}}
\put(000,059){\makebox(0,0)[l]{$h^0,H^0$}}
\put(039,049){\makebox(0,0)[l]{$A^0$}}
\put(039,066){\makebox(0,0)[l]{$A^0$}}
\put(025,040){\makebox(0,0)[b]{4.}}
\put(050,045){\epsffile[072 415 185 486]{top234.eps}}
\put(069,062){\makebox(0,0)[b]{$\tilde{f}_L$}}
\put(069,052){\makebox(0,0)[t]{$\tilde{f}_L$}}
\put(050,059){\makebox(0,0)[l]{$h^0,H^0$}}
\put(089,049){\makebox(0,0)[l]{$A^0$}}
\put(089,066){\makebox(0,0)[l]{$A^0$}}
\put(075,040){\makebox(0,0)[b]{5.}}
\put(100,045){\epsffile[421 415 534 486]{top234.eps}}
\put(119,062){\makebox(0,0)[b]{$Z^0$}}
\put(119,052){\makebox(0,0)[t]{$Z^0$}}
\put(100,059){\makebox(0,0)[l]{$h^0,H^0$}}
\put(139,049){\makebox(0,0)[l]{$A^0$}}
\put(139,066){\makebox(0,0)[l]{$A^0$}}
\put(125,040){\makebox(0,0)[b]{6.}}
\put(000,005){\epsffile[305 415 418 486]{top234.eps}}
\put(019,022){\makebox(0,0)[b]{$W^+$}}
\put(019,012){\makebox(0,0)[t]{$W^-$}}
\put(000,019){\makebox(0,0)[l]{$h^0,H^0$}}
\put(039,009){\makebox(0,0)[l]{$A^0$}}
\put(039,026){\makebox(0,0)[l]{$A^0$}}
\put(025,000){\makebox(0,0)[b]{7.}}
\put(050,005){\epsffile[188 299 302 370]{top234.eps}}
\put(062,016){\makebox(0,0)[tr]{$h^0,H^0$}}
\put(074,015){\makebox(0,0)[bl]{$A^0,G^0$}}
\put(050,023){\makebox(0,0)[l]{$h^0,H^0$}}
\put(089,009){\makebox(0,0)[l]{$A^0$}}
\put(089,026){\makebox(0,0)[l]{$A^0$}}
\put(075,000){\makebox(0,0)[b]{8.}}
\put(100,005){\epsffile[072 299 185 370]{top234.eps}}
\put(112,016){\makebox(0,0)[tr]{$H^+$}}
\put(124,015){\makebox(0,0)[bl]{$G^+$}}
\put(100,023){\makebox(0,0)[l]{$h^0,H^0$}}
\put(139,009){\makebox(0,0)[l]{$A^0$}}
\put(139,026){\makebox(0,0)[l]{$A^0$}}
\put(125,000){\makebox(0,0)[b]{9.}}
\end{picture} 
\caption{vertex corrections, part 1}
\label{vertexdiag1}
\end{figure} 
\epsfclipoff

\epsfclipon
\unitlength1mm
\begin{figure}[htbp]
\begin{picture}(150,200)
%\put(0,0){\framebox(150,210)[t]{}}
\put(000,165){\epsffile[188 298 302 370]{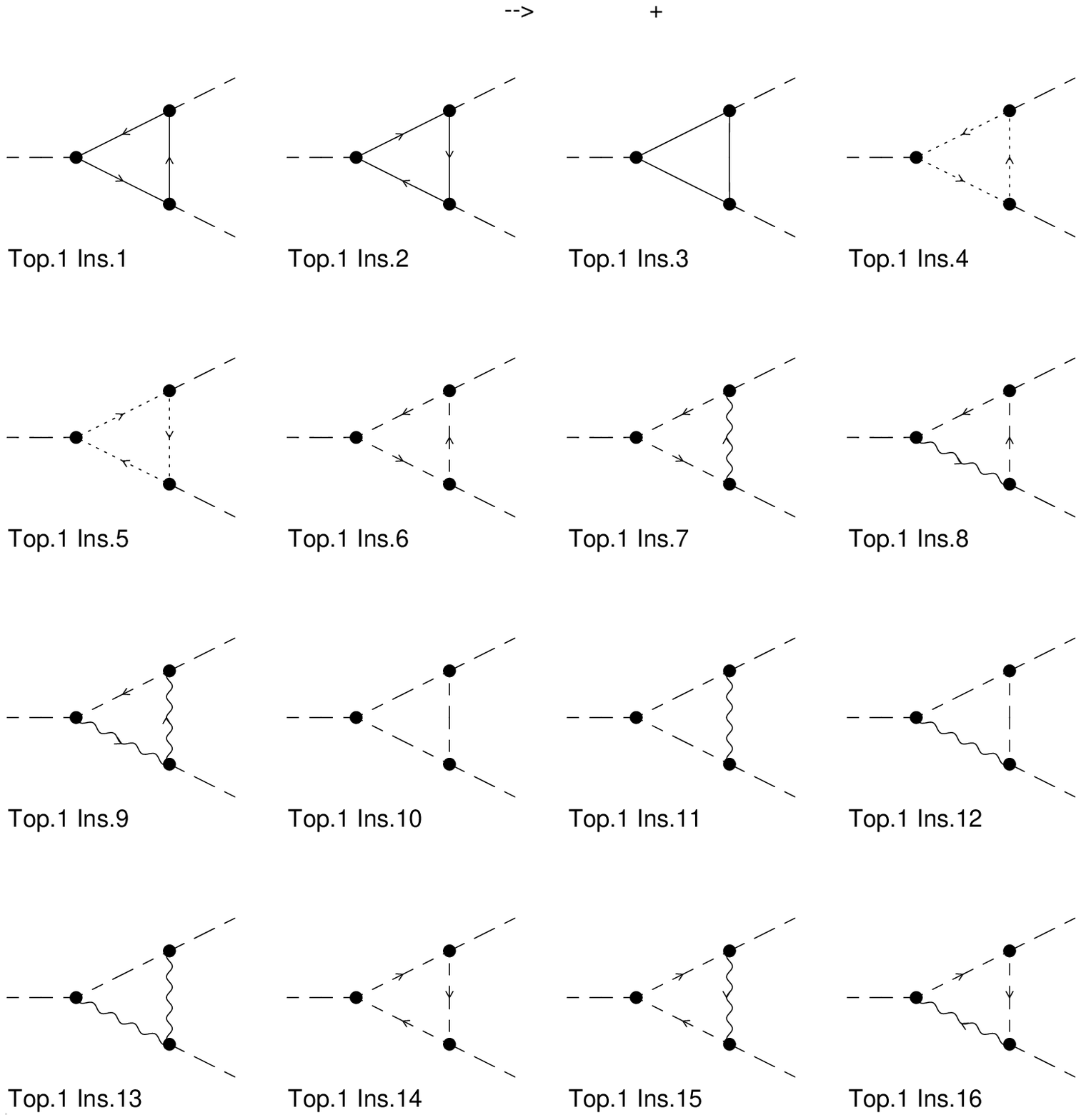}}
\put(019,182){\makebox(0,0)[br]{$h^0,H^0$}}
\put(019,172){\makebox(0,0)[tr]{$h^0,H^0$}}
\put(028,177){\makebox(0,0)[l]{$A^0,G^0$}}
\put(000,179){\makebox(0,0)[l]{$h^0,H^0$}}
\put(037,189){\makebox(0,0)[lt]{$A^0$}}
\put(037,166){\makebox(0,0)[lb]{$A^0$}}
\put(025,160){\makebox(0,0)[b]{10.}}
\put(050,165){\epsffile[188 183 302 253]{top1a.eps}}
\put(069,182){\makebox(0,0)[br]{$H^+$}}
\put(069,172){\makebox(0,0)[tr]{$H^+$}}
\put(078,177){\makebox(0,0)[l]{$G^+$}}
\put(050,179){\makebox(0,0)[l]{$h^0,H^0$}}
\put(087,189){\makebox(0,0)[lt]{$A^0$}}
\put(087,166){\makebox(0,0)[lb]{$A^0$}}
\put(075,160){\makebox(0,0)[b]{11.}}
\put(100,165){\epsffile[188 183 302 253]{top1a.eps}}
\put(119,182){\makebox(0,0)[br]{$G^+$}}
\put(119,172){\makebox(0,0)[tr]{$G^+$}}
\put(128,177){\makebox(0,0)[l]{$H^+$}}
\put(100,179){\makebox(0,0)[l]{$h^0,H^0$}}
\put(137,189){\makebox(0,0)[lt]{$A^0$}}
\put(137,166){\makebox(0,0)[lb]{$A^0$}}
\put(125,160){\makebox(0,0)[b]{12.}}
\put(000,125){\epsffile[188 183 302 253]{top1a.eps}}
\put(019,142){\makebox(0,0)[br]{$\tilde{f}_R$}}
\put(019,132){\makebox(0,0)[tr]{$\tilde{f}_R$}}
\put(028,137){\makebox(0,0)[l]{$\tilde{f}_L$}}
\put(000,139){\makebox(0,0)[l]{$h^0,H^0$}}
\put(037,149){\makebox(0,0)[lt]{$A^0$}}
\put(037,126){\makebox(0,0)[lb]{$A^0$}}
\put(025,120){\makebox(0,0)[b]{13.}}
\put(050,125){\epsffile[188 183 302 253]{top1a.eps}}
\put(069,142){\makebox(0,0)[br]{$\tilde{f}_L$}}
\put(069,132){\makebox(0,0)[tr]{$\tilde{f}_L$}}
\put(078,137){\makebox(0,0)[l]{$\tilde{f}_R$}}
\put(050,139){\makebox(0,0)[l]{$h^0,H^0$}}
\put(087,149){\makebox(0,0)[lt]{$A^0$}}
\put(087,126){\makebox(0,0)[lb]{$A^0$}}
\put(075,120){\makebox(0,0)[b]{14.}}
\put(100,125){\epsffile[188 534 302 605]{top1a.eps}}
\put(119,142){\makebox(0,0)[br]{$f$}}
\put(119,130){\makebox(0,0)[tr]{$f$}}
\put(128,136){\makebox(0,0)[l]{$f$}}
\put(100,138){\makebox(0,0)[l]{$h^0,H^0$}}
\put(137,149){\makebox(0,0)[lt]{$A^0$}}
\put(137,126){\makebox(0,0)[lb]{$A^0$}}
\put(125,120){\makebox(0,0)[b]{15.}}
\put(000,085){\epsffile[188 534 302 605]{top1a.eps}}
\put(018,101){\makebox(0,0)[br]{$\tilde{\chi}_i^+$}}
\put(018,091){\makebox(0,0)[tr]{$\tilde{\chi}_j^+$}}
\put(028,096){\makebox(0,0)[l]{$\tilde{\chi}_k^+$}}
\put(000,098){\makebox(0,0)[l]{$h^0,H^0$}}
\put(037,109){\makebox(0,0)[lt]{$A^0$}}
\put(037,086){\makebox(0,0)[lb]{$A^0$}}
\put(025,080){\makebox(0,0)[b]{16.}}
\put(050,085){\epsffile[305 534 418 605]{top1a.eps}}
\put(068,101){\makebox(0,0)[br]{$\tilde{\chi}_i^0$}}
\put(068,091){\makebox(0,0)[tr]{$\tilde{\chi}_j^0$}}
\put(078,096){\makebox(0,0)[l]{$\tilde{\chi}_k^0$}}
\put(050,098){\makebox(0,0)[l]{$h^0,H^0$}}
\put(087,109){\makebox(0,0)[lt]{$A^0$}}
\put(087,086){\makebox(0,0)[lb]{$A^0$}}
\put(075,080){\makebox(0,0)[b]{17.}}
\put(100,085){\epsffile[188 418 302 489]{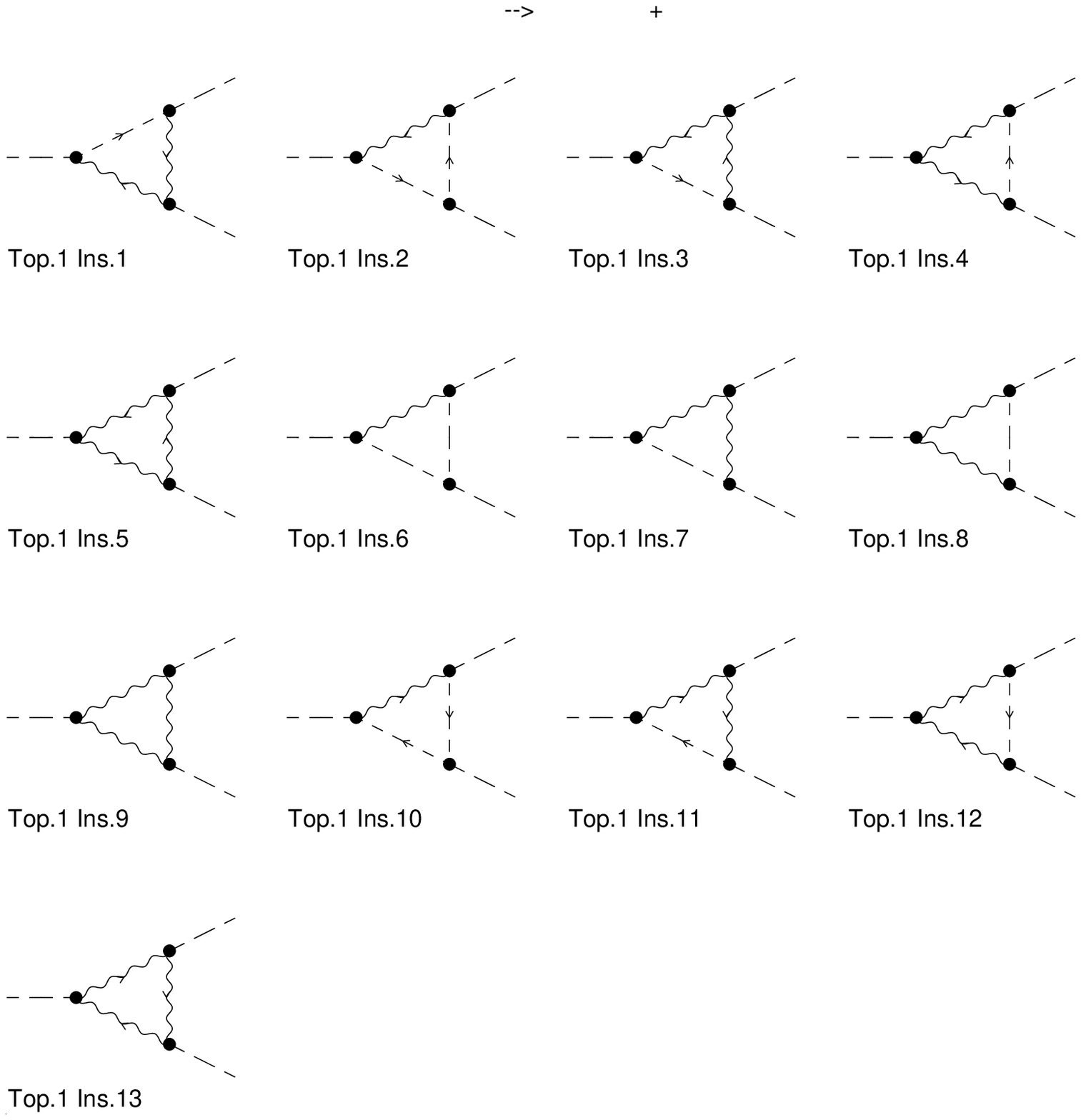}}
\put(119,101){\makebox(0,0)[br]{$Z^0$}}
\put(119,091){\makebox(0,0)[tr]{$A^0,G^0$}}
\put(128,096){\makebox(0,0)[l]{$h^0,H^0$}}
\put(100,098){\makebox(0,0)[l]{$h^0,H^0$}}
\put(137,109){\makebox(0,0)[lt]{$A^0$}}
\put(137,086){\makebox(0,0)[lb]{$A^0$}}
\put(125,080){\makebox(0,0)[b]{18.}}
\put(000,045){\epsffile[188 298 302 370]{top1b.eps}}
\put(019,062){\makebox(0,0)[br]{$W^+$}}
\put(019,052){\makebox(0,0)[tr]{$G^+$}}
\put(028,057){\makebox(0,0)[l]{$H^+$}}
\put(000,059){\makebox(0,0)[l]{$h^0,H^0$}}
\put(037,069){\makebox(0,0)[lt]{$A^0$}}
\put(037,046){\makebox(0,0)[lb]{$A^0$}}
\put(025,040){\makebox(0,0)[b]{19.}}
\put(050,045){\epsffile[305 298 418 370]{top1a.eps}}
\put(069,062){\makebox(0,0)[br]{$h^0,H^0$}}
\put(069,052){\makebox(0,0)[tr]{$h^0,H^0$}}
\put(078,057){\makebox(0,0)[l]{$Z^0$}}
\put(050,059){\makebox(0,0)[l]{$h^0,H^0$}}
\put(087,069){\makebox(0,0)[lt]{$A^0$}}
\put(087,046){\makebox(0,0)[lb]{$A^0$}}
\put(075,040){\makebox(0,0)[b]{20.}}
\put(100,045){\epsffile[305 183 418 253]{top1a.eps}}
\put(119,062){\makebox(0,0)[br]{$H^+$}}
\put(119,052){\makebox(0,0)[tr]{$H^+$}}
\put(128,057){\makebox(0,0)[l]{$W^+$}}
\put(100,059){\makebox(0,0)[l]{$h^0,H^0$}}
\put(137,069){\makebox(0,0)[lt]{$A^0$}}
\put(137,046){\makebox(0,0)[lb]{$A^0$}}
\put(125,040){\makebox(0,0)[b]{21.}}
\put(000,005){\epsffile[421 418 534 489]{top1b.eps}}
\put(019,020){\makebox(0,0)[br]{$Z^0$}}
\put(019,010){\makebox(0,0)[tr]{$Z^0$}}
\put(028,016){\makebox(0,0)[l]{$h^0,H^0$}}
\put(000,017){\makebox(0,0)[l]{$h^0,H^0$}}
\put(037,028){\makebox(0,0)[lt]{$A^0$}}
\put(037,005){\makebox(0,0)[lb]{$A^0$}}
\put(025,000){\makebox(0,0)[b]{22.}}
\put(050,005){\epsffile[421 298 534 370]{top1b.eps}}
\put(069,022){\makebox(0,0)[br]{$W^+$}}
\put(069,012){\makebox(0,0)[tr]{$W^+$}}
\put(078,017){\makebox(0,0)[l]{$H^+$}}
\put(050,019){\makebox(0,0)[l]{$h^0,H^0$}}
\put(087,029){\makebox(0,0)[lt]{$A^0$}}
\put(087,006){\makebox(0,0)[lb]{$A^0$}}
\put(075,000){\makebox(0,0)[b]{23.}}
\put(100,005){\epsffile[188 298 302 370]{top1a.eps}}
\put(119,022){\makebox(0,0)[br]{$A^0,G^0$}}
\put(119,012){\makebox(0,0)[tr]{$A^0,G^0$}}
\put(128,017){\makebox(0,0)[l]{$h^0,H^0$}}
\put(100,019){\makebox(0,0)[l]{$h^0,H^0$}}
\put(137,029){\makebox(0,0)[lt]{$A^0$}}
\put(137,006){\makebox(0,0)[lb]{$A^0$}}
\put(125,000){\makebox(0,0)[b]{24.}}
\end{picture}
\caption{vertex corrections, part 2}
\label{vertexdiag2}
\end{figure}
\epsfclipoff

\noindent
The counter term derived  from (\ref{haapot}) reads in the scheme 
of \cite{ds}:
\BEA
CT_{h^0} = - \frac{ig}{2\cw} \Bigg[ &-& \Sigma_A'(\MA^2)M_Z\KL \frac{3}{2} 
                                                    \Sab\CZb \KR \\ \nonumber
    &+& \re \Sigma_{AZ}(\MA^2)\KL \Cba - \frac{\CQZb\Sab}{\SZb}\KR \\ \nonumber
    &+& \Sigma_\gamma'(0)\frac{M_Z}{2}\CZb\Sab \\ \nonumber
    &+& \re \Sigma_Z(M_Z^2) \frac{1}{M_Z}\KL \CZb\Sab - \frac{\cw^2}{2\sw^2}
                                       \CZb\Sab\KR \\ \nonumber
    &+& \re \Sigma_W(M_W^2) \frac{1}{M_Z}\CZb\Sab \KL \frac{1}{2\sw^2} - 
                                       \frac{1}{2\cw^2} \KR \\ \nonumber
    &-& \Sigma_{\gamma Z}(0) \frac{1}{M_Z} \CZb\Sab \frac{\sw}{\cw} \Bigg] .
\EEA
Analogously,  the \HAA counterterm is given by
\BEA
CT_{H^0} = + \frac{ig}{2\cw} \Bigg[ &-& \Sigma_A'(\MA^2)M_Z\KL \frac{3}{2} 
                                                    \Cab\CZb \KR \\ \nonumber
    &+& \re \Sigma_{AZ}(\MA^2)\KL \Sba - \frac{\CQZb\Cab}{\SZb}\KR \\ \nonumber
    &+& \Sigma_\gamma'(0)\frac{M_Z}{2}\CZb\Cab \\ \nonumber
    &+& \re \Sigma_Z(M_Z^2) \frac{1}{M_Z}\KL \CZb\Cab - \frac{\cw^2}{2\sw^2}
                                       \CZb\Cab\KR \\ \nonumber
    &+& \re \Sigma_W(M_W^2) \frac{1}{M_Z}\CZb\Cab \KL \frac{1}{2\sw^2} - 
                                       \frac{1}{2\cw^2} \KR \\ \nonumber
    &-& \Sigma_{\gamma Z}(0) \frac{1}{M_Z} \CZb\Cab \frac{\sw}{\cw} \Bigg] .
\EEA

\noindent
$\Sigma_a$, $\Sigma_{ab}$ denote the diagonal and non-diagonal self energies,
and $\Sigma '$ the derivative with respect to 
the momentum squared.
As has been checked analytically, the
sum of the vertex diagrams and the counterterm is UV-finite.

\bigskip
\subsection{Approximations}
\begin{figure}[htb]
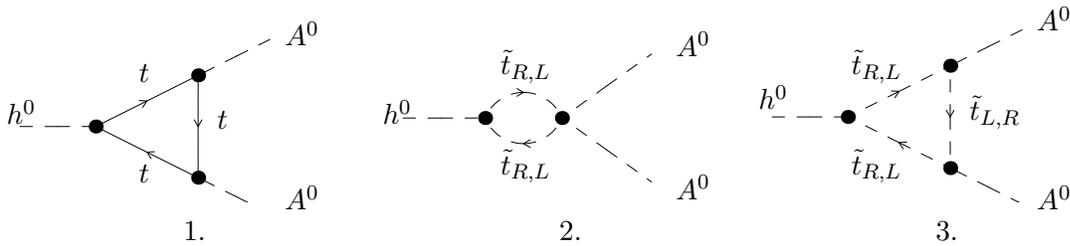

\begin{center}
\begin{picture}(150,40)
%\put(0,0){\framebox(150,040)[t]{}}
\put(000,005){\epsffile[188 534 302 605]{top1a.eps}}
\put(019,021){\makebox(0,0)[br]{$t$}}
\put(019,010){\makebox(0,0)[tr]{$t$}}
\put(028,016){\makebox(0,0)[l]{$t$}}
\put(000,017){\makebox(0,0)[l]{$h^0$}}
\put(037,029){\makebox(0,0)[lt]{$A^0$}}
\put(037,004){\makebox(0,0)[lb]{$A^0$}}
\put(025,000){\makebox(0,0)[b]{1.}}
\put(050,005){\epsffile[072 415 185 486]{top234.eps}}
\put(069,022){\makebox(0,0)[b]{$\tilde{t}_{R,L}$}}
\put(069,012){\makebox(0,0)[t]{$\tilde{t}_{R,L}$}}
\put(050,017){\makebox(0,0)[l]{$h^0$}}
\put(089,007){\makebox(0,0)[l]{$A^0$}}
\put(089,026){\makebox(0,0)[l]{$A^0$}}
\put(075,000){\makebox(0,0)[b]{2.}}
\put(100,005){\epsffile[188 183 302 253]{top1a.eps}}
\put(119,022){\makebox(0,0)[br]{$\tilde{t}_{R,L}$}}
\put(119,012){\makebox(0,0)[tr]{$\tilde{t}_{R,L}$}}
\put(128,017){\makebox(0,0)[l]{$\tilde{t}_{L,R}$}}
\put(100,019){\makebox(0,0)[l]{$h^0$}}
\put(137,031){\makebox(0,0)[lt]{$A^0$}}
\put(137,004){\makebox(0,0)[lb]{$A^0$}}
\put(125,000){\makebox(0,0)[b]{3.}}
\end{picture}
\caption{\label{leaddiag}The leading $h^0A^0A^0$ vertex diagrams}
%\caption{The leading diagrams}
\end{center}
%\label{leaddiag}
\end{figure}

\noindent
For practical purposes it is convenient to consider in particular the subset
of vertex correction diagrams containing terms
$\sim \mt^4$. There are three diagrams of this type, shown in 
Figure~\ref{leaddiag}.
\epsfclipon
\unitlength1mm
\epsfclipoff

\noindent
The coefficient of the $\sim \mt^4$  term in the sum of these
diagrams is UV-finite.
In the approximation $\mt, \mtr, \mtl \gg \Mh, \MA$ one obtains:
\BEA
\label{tnn0}
\Delta T_{h^0, top} = \frac{3g^3\Ca}{4M_W^3 \Sb (4\pi)^2} &\Big[&
            \CTQb \Big( 2 \mt^4(\Mh^2 -2\MA^2)\frac{1}{\mt^2} \\ \nonumber
         & &\qquad\quad +4\mt^4 \ln \big(\frac{\mt^2}
                  {\sqrt{\mtl^2 \mtr^2}}\big) \Big) \\ \nonumber
         & &+2(\mu -A_t \CTb)^2 \mt^4 \frac{\ln\mtr^2 - \ln\mtl^2}
            {\mtl^2-\mtr^2} \Big] .
\EEA

\noindent
The further simplification $\mtl \approx \mtr$ yields
\BEA
\label{tnn}
\Delta T_{h^0, top} = \frac{3g^3\Ca}{4M_W^3 \Sb (4\pi)^2} &\Big[&
        \CTQb \Big( 2 \mt^4(\Mh^2 -2\MA^2)\frac{1}{\mt^2}
                        \quad (I) \hspace{6mm} \\ \nonumber
            & &\qquad \quad +4\mt^4 \ln \frac{\mt^2}{\mts^2}
            \Big) \qquad \hspace*{12mm} (II) \hspace{1mm}\\ \nonumber
            & &-2(\mu -A_t \CTb)^2 \frac{\mt^4}{\mts^2} \Big] 
                  \qquad \hspace*{10mm} (III) \hspace{8mm}.
\EEA

\begin{figure}[htb]
\epsfxsize=15cm
\begin{picture}(150,120)
\put(0,0){\epsffile[0 40 567 500]{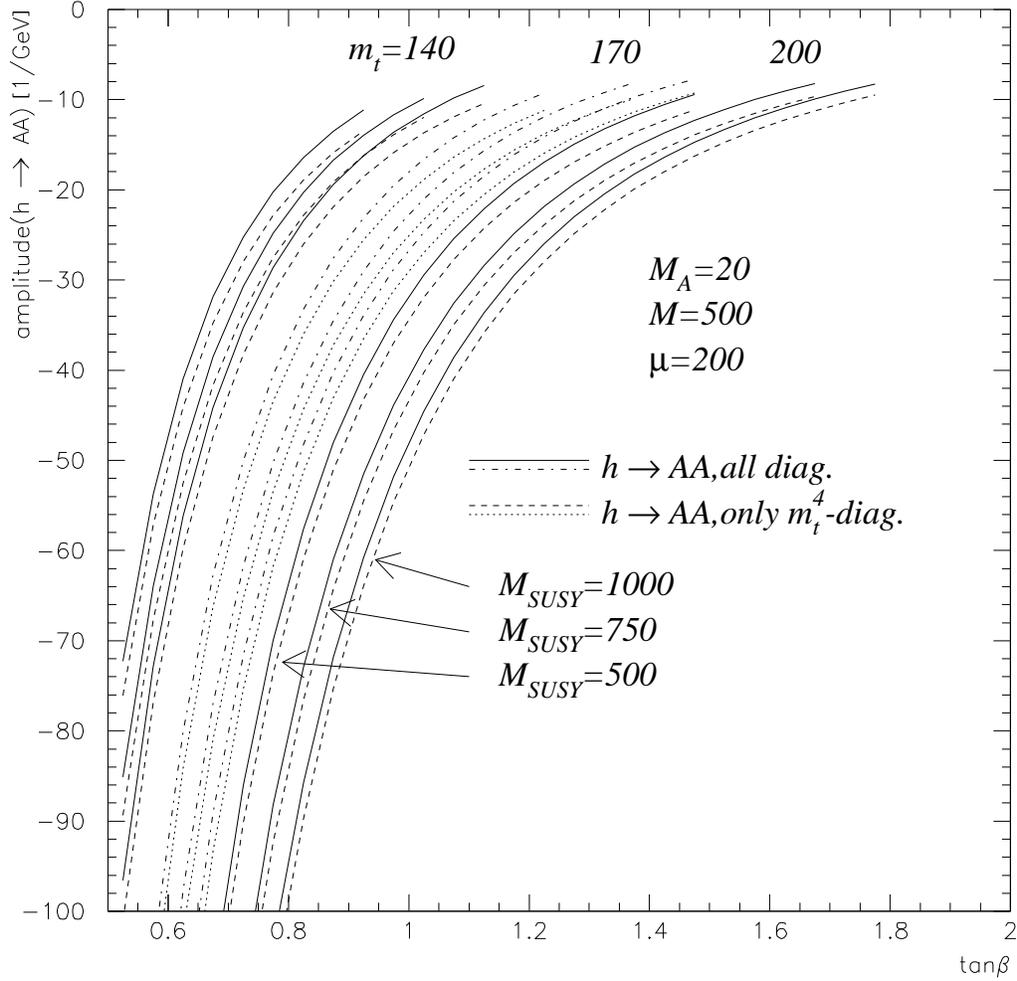}}
\end{picture}
\caption{The \hAA vertex corrections $\Delta T_{h^0}$ and the
            approximation (3.7)}
\label{topnaeh}
\end{figure}

\noindent
This very simple formula approximates the full 1-loop result 
surprisingly well.
The size of the vertex corrections and the quality of the 
approximation (\ref{tnn}) are displayed in Figure~\ref{topnaeh}.

\epsfclipon
\unitlength1mm
\epsfclipoff

\smallskip 
Term $(II)$ reproduces
exactly the result given in~\cite{brignole} for the case $A_t=\mu=0$,
 as well as  the
 $\mt^4$-term of the expression in~\cite{haa}.
In the general case for $A_t,\mu \sim {\cal O}(\mts)$, the
contribution $(III)$  can be
of similar size.
 
\bigskip
%\subsection{Mixed sfermions}
In the case of mixed sfermions, the left and right-handed $\tilde{t}_{L,R}$
have to be replaced by the mass eigenstates of the sfermion mass matrix 
(A.2):   
\BE
\tilde{t}_L, \tilde{t}_R \to \tilde{t}_1, \tilde{t}_2 .
\EE
Each vertex bilinear in the left- and/or right-handed sfermions
 is replaced by four
vertices involving $\tilde{t}_1, \tilde{t}_2$ (see app. A1):
\BE
V(\tilde{t}_{\alpha}, \tilde{t}_{\beta}) =
 \sum_{i,j = 1,2}
T_{\alpha i} T_{\beta j} V(\tilde{t}_i\tilde{t}_j)
\;\;\;\;
 (\alpha, \beta = L, R) .
\EE
The diagrams of Figure~\ref{leaddiag}
 have been computed also for mixed
sfermions. 
The result can be cast into the same form as (\ref{tnn0}) by
substituting $m_{\tilde{t}_{L,R}} \to m_{\tilde{t}_{1,2}}$, the eigenvalues
of the nondiagonal sfermion mass matrix.

\section{The decay ${\mbox h}^0 \to {\mbox A}^0 {\mbox A}^0$}

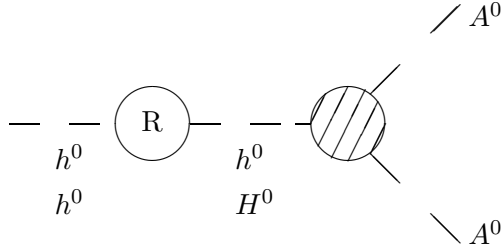
\begin{figure}[htbp]
\begin{center}
\begin{picture}(7,4)
\thinlines
\dline{0.2}{2}{0.8}{0}{2}{1}{0}{0.4}
\dline{2.6}{2}{0.8}{0}{2}{1}{0}{0.4}
\put(4.0,2){\line(1,0){0.2}}
\put(2.1,2){\circle{1.}}
\put(4.7,2){\circle{1.}}
\dline{5}{1.6}{0.8}{-0.8}{2}{1}{-1}{0.39}
\dline{5}{2.4}{0.8}{0.8}{2}{1}{1}{0.39}
\put(4.48,1.55){\line(1,2){0.45}}
\put(4.7,1.5){\line(1,2){0.4}}
\put(4.3,1.7){\line(1,2){0.4}}
\put(4.2,2.0){\line(1,2){0.20}}
\put(5.0,1.6){\line(1,2){0.20}}
\put(0.8,1.4){$h^0$}
\put(0.8,0.8){$h^0$}
\put(3.2,1.4){$h^0$}
\put(3.2,0.8){$H^0$}
\put(1.95,1.9){R}
\put(6.3,0.4){$A^0$}
\put(6.3,3.3){$A^0$}
\end{picture}
\caption{\hAA amplitude with higher order contributions}
\label{einloopamp}
\end{center}
\end{figure}

%\subsection{The amplitude for ${\mbox h}^0 \to {\mbox A}^0 {\mbox A}^0$}
\noindent
The one-loop decay 
 amplitude for $h^0 \to A^0A^0$, graphically depicted in 
Figure~\ref{einloopamp}, can be written as follows:

\unitlength1cm

\BEA
\label{einloopampt1}
T_1 & = &
 \sqrt{Z_{h^0}}\left[  -\frac{gM_Z}{2\cw}\CZb\Sab + \Delta T_{h^0} \right. 
                 \nonumber \\
     &   &  \left. +   Z_{h^0H^0}\KL \frac{gM_Z}{2\cw}\CZb\Cab
               +\Delta T_{H^0} \KR  \right] .
\EEA

\noindent 
$\Delta T_{h,H}$ are the vertex corrections 
(\ref{deltatkh}) and (\ref{deltatgh}). 
 $Z_{h^0}$ is the residue of the $h^0$-propagator in    
(\ref{propmatrix}), which determines the wave-function renormalisation
of the external $h^0$:
\BE
\label{wavefuncren}
Z_{h^0} = \mbox{Res}_{M_{H^0}} \Delta_{h^0} = \frac{1}{1 + \hat{\Sigma}_{h^0}'
(k^2) - \KL \frac{\hat{\Sigma}_{h^0H^0}^2(k^2)}{k^2 - m_{H^0}^2 + 
\hat{\Sigma}_{H^0}(k^2)} \KR '} \Big|_{k^2=M_{h^0}^2}
\EE
The mixing between $h^0$ and $H^0$ enters via
\BE
Z_{h^0H^0} = -\frac{\hat{\Sigma}_{h^0H^0}(M_{h^0}^2)}{M_{h^0}^2 - m_{H^0}^2 +
             \hat{\Sigma}_{H^0}(M_{h^0}^2)}.
\EE

%In formula (\ref{einloopampt1}) the full one-loop $H^0A^0A^0$-vertex
%has been inserted. The importance of this operation is shown in 
%fig.~\ref{fampvsapp} . Here ``full amplitude'' means formula 
%\ref{einloopampt1}, ``approximation'' means the same but leaving out
%all \HAA-vertex corrections, i.e. $i\Delta T_{H^0}$.

%\epsfclipon
%\unitlength1mm
%\begin{figure}[htb]
%\epsfxsize=15cm
%\begin{picture}(150,120)
%\put(0,0){\epsffile[0 40 567 500]{khav1v11v.eps}}
%\end{picture}
%\caption{Importance of \HAA-vertex corrections}
%\label{fampvsapp}
%\end{figure}
%\epsfclipoff

\bigskip
%\subsection{$\mt^4$-approximation}
\noindent
A simple  approximation for the full amplitude includes only the
leading term $\omt$ in eq.\ (2.13).
%Higgs bosons in the $\mt^4$ approximation $M_{h^0,\omt}$ 
%(\ref{omegatHiggsmassen}).
The mass  matrix  
\BE
M^2_{Higgs}=\frac{\SZb}{2}\ML \CTb M_Z^2 + \Tb \MA^2 & -M_Z^2 -\MA^2 \\
                            -M_Z^2 -M_A^2 & \Tb M_Z^2 + \CTb \MA^2 + \omt \MR
\EE
can be diagonalised by $\aeff$ \cite{ds}, which is determined by 
\BE
\label{taeff}
\tan \aeff = \frac{-(\MA^2 + M_Z^2)\Tb}
                  {M_Z^2+\MA^2\TQb-(1+\TQb) M_{h^0,\omt}}.
\EE

\noindent
$\aeff$ replaces the lowest order quantity in the tree level vertex
(\ref{bornvertex}). 
The use of $\aeff$ is equivalent to the diagrammatic inclusion of the $h^0H^0$
mixing term
 with respect to the $m_{top}^4$ content.
The corresponding  approximation for the one-loop amplitude is
obtained by
\BE
\label{einloopamtaeff}
T_{1,\alpha_{eff}}  = 
   \left[ -\frac{gM_Z}{2\cw}\CZb\Sab + \Delta T_{h^0} \right]_
   {\alpha \to \aeff} \nonumber \\
\EE

\begin{figure}[htb]
\epsfxsize=19cm
\begin{picture}(190,150)
\put(0,0){\epsffile[2 40 567 500]{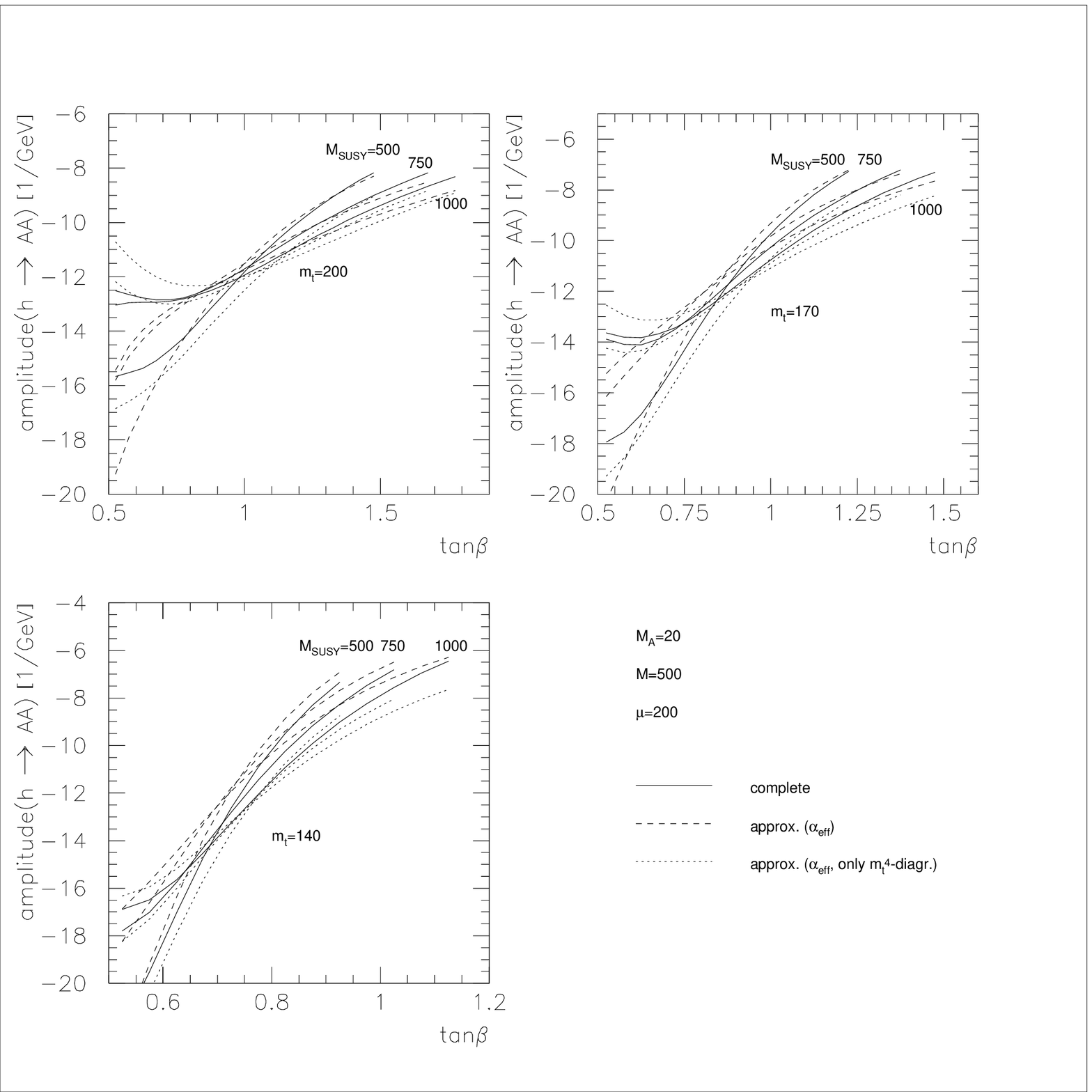}}
\end{picture}
\caption{Full 1-loop amplitude $T_1$ in (\ref{einloopampt1}) 
         and the approximations with $\aeff$ in (\ref{einloopamtaeff}).}
\label{aeffcomp}
\end{figure}

\noindent
In the further simplification of this expression,
$\Delta T_{h^0}$
 is replaced by $\Delta T_{h^0,top}$ of Eq.\ (3.7).
The comparison of the full one-loop amplitude and the two
approximations is shown in Figure~\ref{aeffcomp}.

\epsfclipon
\unitlength1mm
\epsfclipoff

\bigskip \noindent
The decay width for
 \htAA 
\BE
\label{zerfbr}
\Gamma_1(h^0 \to A^0A^0) = \frac{1}{32 \pi \Mh}\sqrt{1-4\frac{\MA^2}{\Mh^2}}
                           \hspace{2mm} |T_1|^2
\EE
is determined at the 1-loop level by the mass $M_{h^0}$ and the decay
amplitude  $T_1$ in (\ref{einloopampt1}).

\bigskip \noindent
%\subsection{Branching ratios}
Of practical importance are
the branching ratios 
\BE
\frac{\Gamma_1(h^0 \to A^0A^0)}{\Gamma_{tot}}, \quad
\frac{\Gamma_1(h^0 \to b\bar{b})}{\Gamma_{tot}}
%\frac{\Gamma_1(h^0 \to c\bar{c})}{\Gamma_{tot}}, \quad
%\frac{\Gamma_1(h^0 \to \tau^+\tau^-)}{\Gamma_{tot}}, \quad
%\frac{\Gamma_1(h^0 \to gg)}{\Gamma_{tot}}
\EE
  where the total width is calculated as
\BE
\Gamma_{tot} = \Gamma_1(h^0 \to A^0A^0) +
\Gamma_1(h^0 \to b\bar{b}) + \Gamma_1(h^0 \to c\bar{c}) + 
\Gamma_1(h^0 \to \tau^+\tau^-) + \Gamma_1(h^0 \to gg).
\EE
The decay $h^0 \to \tilde{\chi}^+_i\tilde{\chi}^-_j$
does not occur because such light charginos are excluded from the
direct search. This also 
suppresses the decay $h^0 \to \tilde{\chi}^0_i\tilde{\chi}^0_j$.

\smallskip
The decay widths for  
 $h^0 \to f\bar{f}$ have been calculated taking into account the full set 
of propagator corrections given in ref.~\cite{ds} and the QCD
corrections given in ref.~\cite{djouadi}.
The full set of electroweak vertex corrections is also available \cite{ds1}, 
but they are numerically of subleading size and can be neglected at
a sufficient level of accuracy.
The decay widths for $h^0 \to gg$ has been computed taking into
account the top- and the bottom-loop, as pointed out in
ref.~\cite{djouadi}.

\section{Discussion}
\begin{figure}[htb]
\epsfxsize=15cm
\begin{picture}(150,120)
\put(0,0){\epsffile[40 40 527 500]{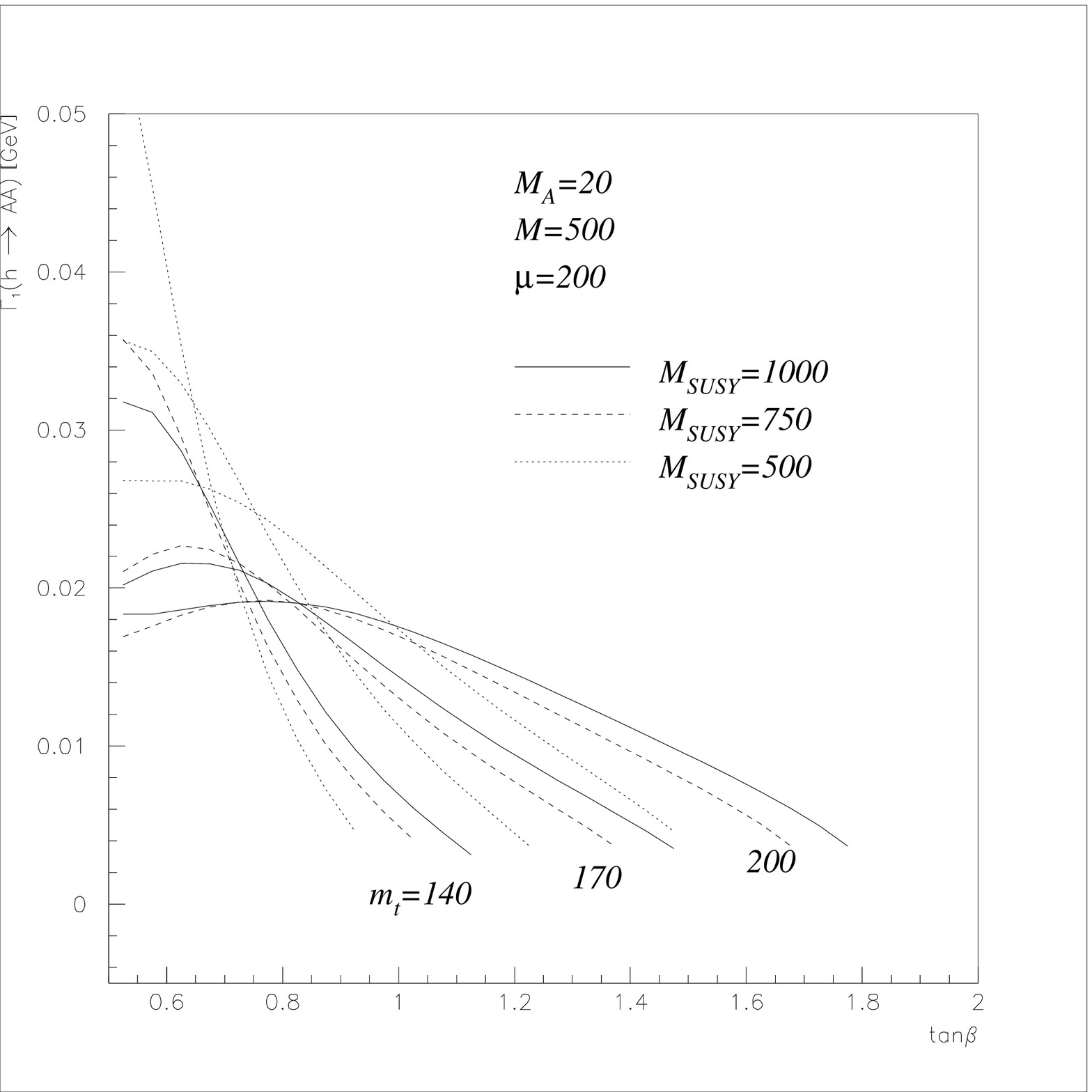}}
\end{picture}
\caption{One-loop decay width $\Gamma_1$(\htAA)}
\label{haazerfbr}
\end{figure}

\noindent
Loop effects are of basic importance for the decay 
\htAA in a twofold respect:
The loop contributions to the mass spectrum make the decay 
kinematically possible, and the vertex corrections give
a sizeable enhancement of the effective \hAA coupling  
over the tree level coupling, which would even vanish for
$\tan\beta=1$.
The partial decay width $\Gamma_1(h^0\rightarrow A^0A^0)$
from the full 1-loop calculation, Eq.~(\ref{zerfbr}), is shown in
Figure \ref{haazerfbr}. It is worth to note that the simplified
approximation based on Eq.~(\ref{einloopamtaeff}) 
with $\aeff$ from (\ref{taeff}) and
the vertex corrections (\ref{tnn0}) yield results which deviate not
more 
than 10\% from the full calculation. The quality of the 
approximations is made more explicit in table~\ref{zerfvergl}, where for
various sets of SUSY parameters the approximate and full 1-loop
results are compared. As a general feature it appears that the
approximations are suitable for the most cases.
Only for unreasonably small values of $\tan\beta$
the approximation becomes unreliable (see Figure \ref{aeffcomp}).

\smallskip 
The dependence on the top mass is much weaker as naively expected
from the expressions in (\ref{tnn0}). The powers of 
$m_t$ are actually
compensated by the $\mt$-dependence of $\cos\aeff$, thus leaving
only a logarithmic behaviour with $\mt$, as pointed out in
ref.~\cite{haa}.  

\begin{figure}[htb]
\epsfxsize=12cm
\epsfysize=10cm
\begin{picture}(120,100)
\put(20,0){\epsffile[40 40 527 500]{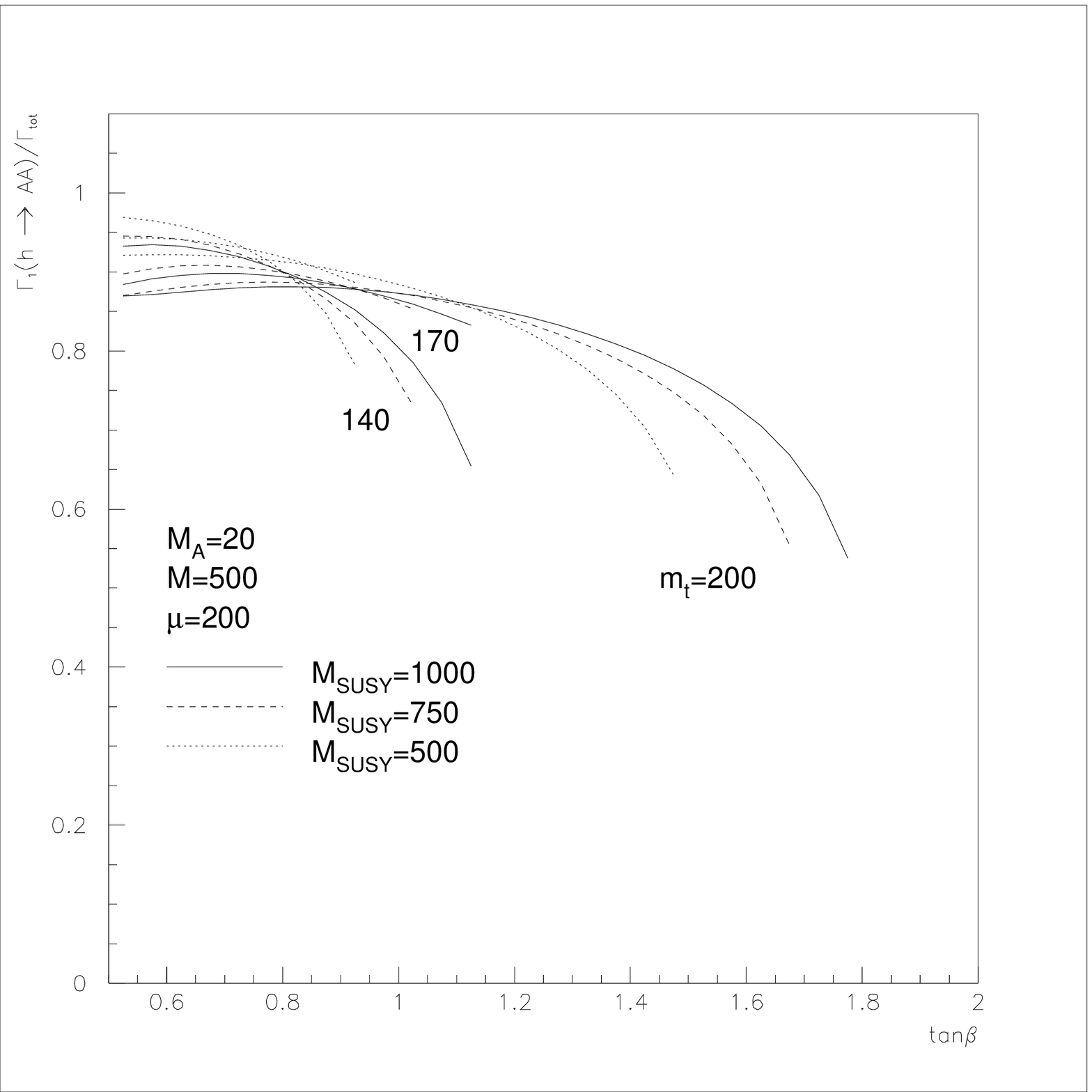}}
\end{picture}
\caption{Branching ratio $\Gamma_1(h^0 \to A^0A^0)/\Gamma_{tot}$}
\label{brhaa}
\end{figure}

\begin{figure}[htb]
\epsfxsize=12cm
\epsfysize=10cm
\begin{picture}(120,100)
\put(20,0){\epsffile[40 40 527 500]{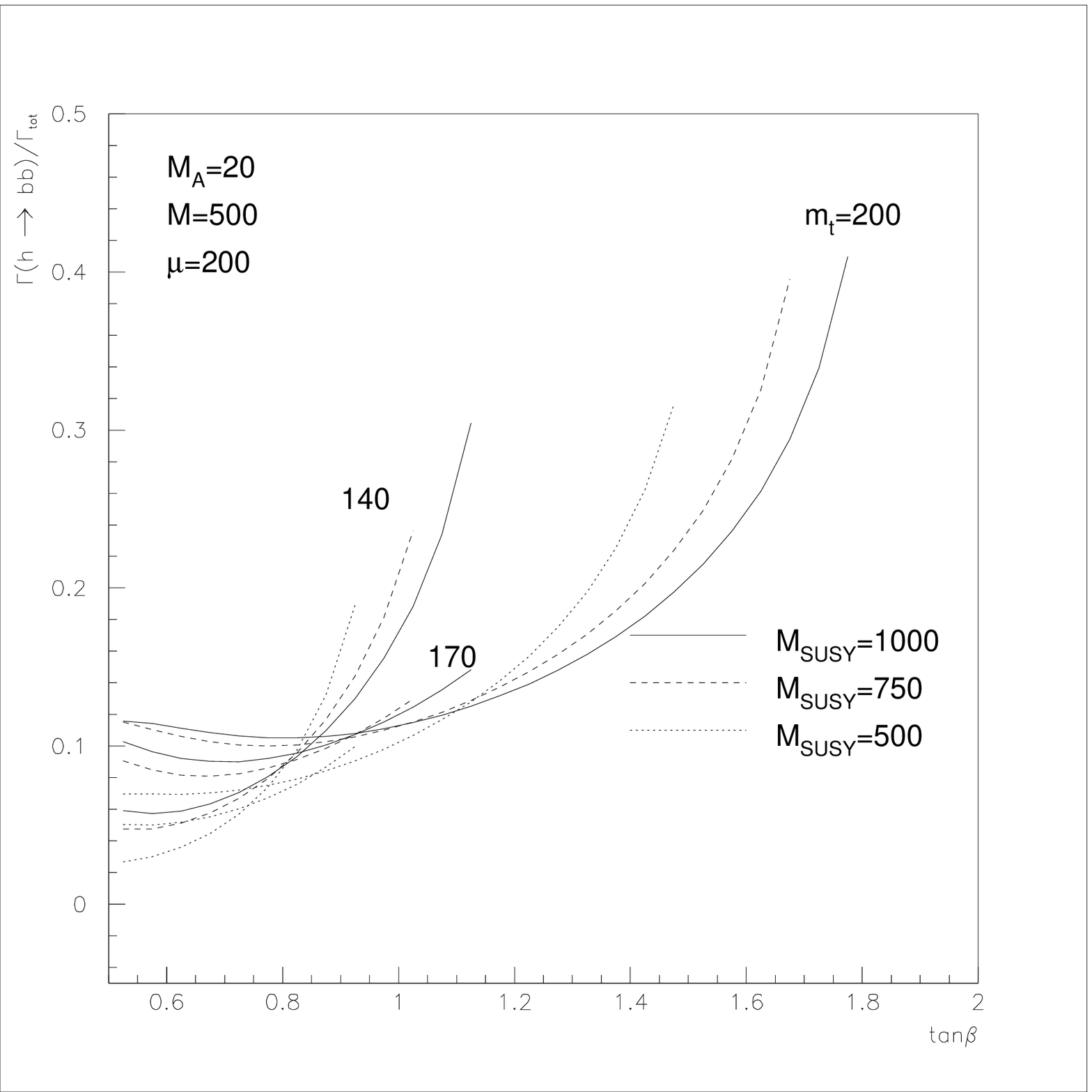}}
\end{picture}
\caption{Branching ratio $\Gamma_1(h^0 \to b\bar{b})/\Gamma_{tot}$}
\label{brhbb}
\end{figure}

\bigskip
The results for the branching ratios 
$$
BR({\mbox A}^0{\mbox A}^0)
 = \frac{\Gamma_1(h^0\rightarrow A^0A^0)}{\Gamma_{tot}}, \;\;\;\;
BR(b\bar{b})
 = \frac{\Gamma_1(h^0\rightarrow b\bar{b})}{\Gamma_{tot}}
$$
are displayed in Figures \ref{brhaa} and \ref{brhbb}.
For practically the entire range where the decay $h^0\rightarrow A^0A^0$
is allowed, it is the dominant decay mode. The branching ratio
$BR(\tau^+\tau^-)$ for decays into $\tau$ pairs does not exceed 6\%.
The decay modes into $c\bar{c}$ and $gg$ are suppressed via the
$\mt$-dependence of $\cos\aeff$. They never exceed 1\% and can thus be
neglected. 

\begin{table}[t]
\begin{center}
\begin{tabular}{|c|c|c|c|} \hline
$\mu$ [GeV] & $\MS$ [MeV] & $\Gamma_1 $ [MeV] & $\Gamma_{1,\aeff}$ [MeV] \\
\hline \hline
 50 & 1000 & 15.9  & 14.6  \\ \hline
300 & 1000 & 17.3  & 17.0  \\ \hline
 50 &  500 & 14.4  & 15.1  \\ \hline
300 &  500 & 29.9  & 31.2  \\ \hline
 50 &  300 & 12.2  & 14.0  \\ \hline
300 &  300 & 88.0  & 90.2  \\ \hline
\end{tabular} \\
\caption{Deacy width $\Gamma_1$, complete at 1-loop, and the approximation
         $\Gamma_{1,\aeff}$ from the amplitude (4.6) with 
         $\Delta T_{h^0,top}$ in (3.7). 
  $\mt=200 GeV, \MA=25 GeV, \Tb=0.8, M=500 GeV$, diagonal sfermion
  mass matrix} 
\label{zerfvergl}
\end{center}
\end{table}

%\noindent

\epsfclipon
\unitlength1mm
\epsfclipoff

\epsfclipon
\unitlength1mm
\epsfclipoff

\epsfclipon
\unitlength1mm
\epsfclipoff

\clearpage

\smallskip \noindent
Our results agree fairly well with those of refs.~\cite{brignole,haa}.
The reason is the already mentioned good quality of an
approximate calculation with terms involving only top and stops.
For comparison, we put together the results of the complete
calculation and of the RG method of ref.~\cite{haa}  in table~\ref{hhnvergl}
in terms of the ratio $r_{h^0A^0A^0}$ of the 1-loop $h^0A^0A^0$ vertex in
Eq.~(3.2), (3.3) and the corresponding expression in ref.~\cite{haa}. 
 Both results are very close for large values of
 $M_{SUSY}$. They deviate more for lower values of the stop masses and
with increasing $\mu, A_t$, as expected from 
the last term of Eq.~(3.8).

\begin{table}[t]
\begin{center}
\begin{tabular}{|c|c|c|} \hline
$\mu$ [GeV] & $\MS$ [GeV] & $r_{h^0A^0A^0}$ \\ \hline \hline
  0 & 1000 & 0.989 \\ \hline
 50 & 1000 & 0.989 \\ \hline
300 & 1000 & 1.042 \\ \hline
  0 &  500 & 1.044 \\ \hline
 50 &  500 & 1.055 \\ \hline
300 &  500 & 1.403 \\ \hline
  0 &  300 & 1.389 \\ \hline
 50 &  300 & 1.443 \\ \hline
300 &  300 & 3.244 \\ \hline
\end{tabular} \\
\end{center}
\caption{The ratio $r_{h^0A^0A^0}$ of the effective \hAA couplings of this
  paper and 
   ref~\protect\cite{haa} (see text). The 
   parameters are as in  table~\ref{zerfvergl}. }
\label{hhnvergl}
\end{table}

We have not implemented the recently calculated leading 2-loop 
results to the neutral MSSM Higgs boson masse \cite{twoloop},
based on effective potential and renormalization group methods.
They would modify the complete 1-loop results in the same way as
the approximations and thus do not influence the differences
discussed here which are only obtained by an explicit diagrammatic
calculation. 

\subsection*{Acknowledgement}

\noindent
We want to thank A.\ Dabelstein for providing us with his routine 
for the one-loop  Higgs masses and propagator corrections,
and A. Djouadi,  
A.\ Kraft and W.\ M\"osle for helpful discussions. 

\newpage \noindent

\vspace{2cm}
\vorn {\Large \bf Appendix}

\begin{appendix}
\section{Masses and constants}

\subsection{Sfermions}

\noindent
The sfermion mass term of the Lagrangian
\BE
{\cal L}_{sferm,mass} = -\frac{1}{2} (\tilde{f}_L,\tilde{f}_R)^\dagger Z 
                        \VL \tilde{f}_L \\ \tilde{f}_R \VR
\EE

\vorn contains the mass matrix
\BE
Z = \ML \tilde{M}_{\tilde{Q}}^2 + M_Z^2\CZb(I_3 - Q_f\sw^2) + m_f^2 & 
       m_f (A_f + \mu \{\cot, \tan\}\beta) \\
       m_f (A_f + \mu \{\cot, \tan\}\beta) &
       \tilde{M}_{\tilde{U},\tilde{D}}^2 + M_Z^2 \CZb Q_f\sw^2 + m_f^2 \VR .
\EE

\vorn  $\{\cot, \tan\}$ refer to the corresponding fermions
with isospin $\{+\frac{1}{2},-\frac{1}{2}\}$.
Unmixed sfermions are obtained by putting
\BE
A_d = -\mu\Tb , \quad A_u = -\mu\CTb
\EE
which, however, is only important for stops  since the non-diagonal terms
are suppressed for the light quarks and low $\tan\beta$.
For the simplest case we assume equal soft breaking 
parameters for all sfermions:
\BE
\label{msusy}
\tilde{M}_{\tilde{Q}} = \tilde{M}_{\tilde{U}} = \tilde{M}_{\tilde{D}} =:\MS
\EE

\noindent
For mixed sfermions
the coefficients of the transformation to the mass eigenstates 
$\tilde{t}_{1,2}$
\BE
\tilde{t}_{\alpha} = \sum_{j=1,2} T_{\alpha j}\, \tilde{t}_j
\;\;\;\;\;\; (\alpha=L,R)
\EE
enter the bilinear stop vertices of the diagrams in section 3,
according to Eq.\ (3.10).
% the conversion coefficients are
%necessary. They can be found in \cite[p. 393]{hhg}.

\subsection{Charginos}
\noindent
The chargino mass term
\BE
{\cal L}_{Char,mass} = -\frac{1}{2}(\psi^+,\psi^-) \ML 0 & X^T \\ X & 0 \MR
                        \VL \psi^+ \\ \psi^- \VR +\hc
\EE

\vorn with
\BE
X = \ML M & M_W\sqrt2 \Sb \\ M_W\sqrt2 \Cb & -\mu \MR,
\EE

%\vorn Die Charginos sind Mischungen aus den Winos und den geladenen Higgsinos:
%\BE
%\psi^\pm = (\tilde{W}^\pm, \tilde{\psi}^\pm_{H_{2,1}})
%\EE

\vorn contains the SU(2) gaugino mass parameter $M$ and  
the Higgsino doublet mixing parameter $\mu$.
The  mass matrix is diagonalized with the help of the unitary
$(2 \times 2)$-matrizes $U$ and $V$:
\BEA
\tilde{\chi}_i^+ &=& V_{ij}\psi_j^+, \\ \nonumber
\tilde{\chi}_i^- &=& U_{ij}\psi_j^-.
\EEA

\vorn $U$ and $V$ are given by 
\BEA \vspace{15mm}
U &=& O_- \\ \nonumber
V &=& \left\{ \begin{array}{cc} O_+ & ,\det X>0 \\ \sigma_3O_+ & ,\det X<0
              \end{array} \right.
\EEA
with
\BE
O_\pm = \ML \cos\phi_\pm & \sin\phi_\pm \\ -\sin\phi_\pm & \cos\phi_\pm \MR,
\quad \sigma_3 = \ML 1 & 0 \\ 0 & -1 \MR.
\EE
$\cos\phi_\pm$ and $\sin\phi_\pm$ are given by
$ ( \epsilon = sgn( \det X)$)
\BEA
\tan \phi_+ &=& \frac{\sqrt2 M_W (\Sb M_+ + \epsilon \Cb M_-)}
                     {M M_+ + \epsilon \mu M_-} \nonumber \\
\tan \phi_- &=& \frac{(-\mu M_+ - \epsilon M M_-)}
                     {\sqrt2 M_W (\Sb M_+ + \epsilon \Cb M_-)}
\EEA

\noindent
If $\phi_+ < 0$ then $\phi_{+} \rightarrow \phi_{+}+\pi$.

\noindent
$M_+$ and $M_-$ are the entries in the diagonalized
chargino mass-matrix
\BE
M_{D,Char} = U^*XV^{-1} = \ML M_+ & 0 \\ 0 & M_- \MR
\EE
obeying
\BEA
M_\pm^2  & = &
 \frac{1}{2} \Big\{ M^2 + \mu^2 + 2M_W^2   \nonumber \\
 &   &   
 \pm \big[ (M^2-\mu^2)^2
   + 4M_W^4\CQZb + 4M_W^2(M^2+\mu^2+2M\mu\SZb) \big]^{\frac{1}{2}} \Big\}.
\EEA

\subsection{Neutralinos}
\noindent
The neutralino mass term
%sind Mischungszust"ande aus dem Photino, dem Zino und den neutralen Higgsinos:
%\BE
%\psi^0 = (\tilde{B}^0,\tilde{W}^3,\psi_{H_1}^0,\psi_{H_2}^0)
%\EE
\BE
{\cal L}_{Neutr,mass} = -\frac{1}{2} (\psi^0)^T Y \psi^0 +\hc .
\EE

\vorn is determined by the matrix 
\BE
Y = \MLv M' & 0 & -M_Z\sw\Cb & M_Z\sw\Sb \\ 0 & M & M_Z\cw\Cb & -M_Z\cw\Sb \\
-M_Z\sw\Cb & M_Z\cw\Cb & 0 & \mu \\ M_Z\sw\Sb & -M_Z\cw\Sb & \mu & 0 \MR
\EE

\vorn The physical neutralino mass-eigenstates can be computed with the help
of a unitary diagonalization matrix $N$:
\BE
\tilde{\chi}_i^0 = N_{ij}\psi_j^0
\EE

\vorn The diagonal mass matrix
is given by:
\BE
M_{D,Neutr} = N^*YN^{-1}
\EE

%In dem nach \ref{charmass} erlaubten Parameterbereich ergeben sich damit
%teilweise Neutralinomassen derart, da"s der Zerfall
%\BE
%h^0 \to \tilde{\chi}^0 \tilde{\chi}^0
%\EE
%
%kinematisch erlaubt ist. Dies gilt jedoch nicht f"ur die in dieser Arbeit
%betrachteten Punkte in der $\mu-M$-Ebene mit der einen Ausnahme von $\mu = 250
%\GeV, M = 100 \GeV$.
\noindent 
The matrix $Y$ contains 
the U(1) gaugino mass $M'$ parameter as a further input quantity.
 As conventionally done, we assume
the GUT constraint
\BE
M' = \frac{5}{3}\frac{\sw^2}{\cw^2}M.
\EE

%Besitzt $Y$ negative Eigenwerte, so werden diese als negative Massen verwendet,
%es sei denn, sie gehen als Argumente in $B_0-$ oder $C_0-$Funktionen ein,
%vgl dazu \cite[S. 388]{hhg}.

\section{Vertex corrections for ${\mbox h}^0 \to {\mbox A}^0{\mbox A}^0$}

\noindent
This section contains the analytical expressions for all one-loop corrections
which contribute to the 
\hAA -vertex. For the Feynman rules  we adopt the 
convention of \cite{hhg}.
The enumeration refers to section 3,
fig.~\ref{vertexdiag1} and \ref{vertexdiag2}.
The indices denote the particles inside the loop with assignment
of masses as follows:

\begin{center}
\unitlength1mm
\begin{picture}(110,50)
\thinlines
\put(0,30){\line(1,0){15.5}}
\put(22.5,30){\circle{15}}
\put(30,30){\line(1,-1){15}}
\put(30,30){\line(1,1){15}}
\put(06,33){\vector(1,0){4}}
\put(07,35){\makebox(0,0)[b]{$h^0$}}
%\put(20,40){\vector(1,0){4}}
%\put(21,42){\makebox(0,0)[b]{$p$}}
%\put(20,20){\vector(1,0){4}}
%\put(21,18){\makebox(0,0)[t]{$k+p$}}
\put(22.5,35){\makebox(0,0)[t]{$m_1$}}
\put(22.5,25){\makebox(0,0)[b]{$m_2$}}
\put(35,32){\vector(1,1){4}}
\put(41,38){\makebox(0,0)[t]{$A^0$}}
\put(39,24){\vector(-1,1){4}}
\put(41,22){\makebox(0,0)[b]{$A^0$}}

\put(60,35){\line(1,0){15.5}}
\put(80.25,29.75){\circle{15}}
\put(85,24.5){\line(1,-1){11}}
\put(75.5,35){\line(1,1){15}}
\put(66,38){\vector(1,0){4}}
\put(67,40){\makebox(0,0)[b]{$h^0$}}
%\put(71,27){\vector(1,-1){4}}
%\put(69,25){\makebox(0,0)[tr]{$p$}}
%\put(90,32){\vector(-1,1){4}}
%\put(92,34){\makebox(0,0)[bl]{$p+q_2$}}
\put(83.5,33){\makebox(0,0)[t]{$m_1$}}
\put(77.5,27){\makebox(0,0)[b]{$m_2$}}
\put(85,42){\vector(1,1){4}}
\put(91,48){\makebox(0,0)[t]{$A^0$}}
\put(94.5,18.5){\vector(-1,1){4}}
\put(96.5,16.5){\makebox(0,0)[b]{$A^0$}}
\end{picture}
\end{center}

\begin{center}
\unitlength1mm
\begin{picture}(060,50)
\thinlines
\put(0,30){\line(1,0){15}}
\put(15,30){\line(2,1){35}}
\put(15,30){\line(2,-1){35}}
\put(35,20){\line(0,1){20}}
\put(06,33){\vector(1,0){4}}
\put(07,35){\makebox(0,0)[b]{$h^0$}}
%\put(38,32){\vector(0,-1){4}}
%\put(40,30){\makebox(0,0)[l]{$p$}}
%\put(23,36){\vector(2,1){4}}
%\put(25,38){\makebox(0,0)[br]{$q_1+p$}}
%\put(27,22){\vector(-2,1){4}}
%\put(25,22){\makebox(0,0)[tr]{$q_2+p$}}
\put(40,40){\vector(2,1){4}}
\put(45,43){\makebox(0,0)[tl]{$A^0$}}
\put(44,18){\vector(-2,1){4}}
\put(45,19){\makebox(0,0)[tl]{$A^0$}}
\put(35,30){\makebox(0,0)[r]{$m_0$}}
\put(25,35){\makebox(0,0)[tl]{$m_1$}}
\put(25,25){\makebox(0,0)[bl]{$m_2$}}
\end{picture}
\end{center}

\noindent
An overall factor $g^3/(4\pi)^2$ is suppressed. The symmetry factors are
included. The color factor $N_f$ which occurs for fermion and sfermion loops
is $3$ for quarks and squarks, $1$ otherwise.
$f$ denotes
 the fermions with weak isospin $+\frac{1}{2}$, 
$f'$ their isospin partners with isospin $-\frac{1}{2}$.
Summation over the various generations is understood.

\smallskip
For given external momenta
the first two arguments of the $C_0$-functions are the same in all diagrams
and  thus have also been dropped:
 $C_0(-q_1,q_2,m_1,m_0,m_2) \equiv
 C_0(m_1,m_0,m_2)$.     Here
$-q_1$ and $q_2$ denote the momenta of the two outgoing $A^0$ and 
$(-q_1+q_2)=k$  denotes the momentum of the incoming $h^0$ 
with $k^2 = M_{h^0}^2$.
In order to get the corresponding results for $\Delta T_{H^0,i}$ one
 has to replace the $h^0$-coupling
by the corresponding $H^0$-coupling according to the
Feynman rules of ref.~\cite{hhg}.

%Bei der in dieser Arbeit verwendeten numerischen $C_0$-Funktion wird, da sie
%nur f"ur Teilchen identischer Masse in den beiden Endzust"anden den Wert
%berechnen kann, nur das Quadrat der Summe der beiden angegebenen Impulse und das
%Quadrat eines der Impulse (welches aus oben genannten Gr"unden mit dem Quadrat
%des anderen Impulses "ubereinstimmen mu"s) verwendet. Da in allen vorkommenden
%$C_0$-Funktionen diese beiden Werte aus $\Mh^2$ und $\MA^2$ bestehen, werden
%diese beiden Argumente weggelassen und nur die noch ben"otigten Massen
%angegeben: $C_0(-q_1,q_2,m_1,m_0,m_2) \equiv C_0(m_1,m_0,m_2)$.

\BEA
\Delta T_{ 1,h^0h^0} &=& \frac{3iM_Z}{2(2\cw)^3}\CZb\CQZa\Sab
                         B_0(\Mh^2,\mh,\mh) \\
\Delta T_{ 1,h^0H^0} &=& \frac{-iM_Z}{(2\cw)^3}\CZb\SZa(2\SZa\Sab-\Cab\CZa)
                         \nonumber \\
                     & & \hspace{7cm} B_0(\Mh^2,\mh,\mH) \\
\Delta T_{ 1,H^0H^0} &=& \frac{iM_Z}{2(2\cw)^3}\CZb\CZa(2\SZa\Cab+\Sab\CZa)
                         \nonumber \\
                     & & \hspace{7cm} B_0(\Mh^2,\mH,\mH) \\
\Delta T_{ 2,A^0A^0} &=& \frac{3iM_Z}{2(2\cw)^3}\CDZb\Sab B_0(\Mh^2,\MA,\MA)\\
\Delta T_{ 2,A^0G^0} &=& \frac{3iM_Z}{(2\cw)^3}\CZb\SQZb\Sab
                         B_0(\Mh^2,\MA,M_Z)\\
\Delta T_{ 2,G^0G^0} &=& \frac{-iM_Z}{2(2\cw)^3}\CZb\Sab (3\SQZb-1)
                         B_0(\Mh^2,M_Z,M_Z)\\
\Delta T_{ 3,H^+H^-} &=& \frac{i}{(2\cw)^3}\CDZb\Sab
                         \KL M_W\Sba+\frac{M_Z}{2\cw}\KR \nonumber \\
                     & & \hspace{7cm} B_0(\Mh^2,\MP,\MP)\\
\Delta T_{ 3,H^+G^-} &=& \frac{2i}{(2\cw)^3}\CZb\SZb
                         (\cw M_W\Cba-M_Z\SZb\Sab)  \nonumber \\
                     & & \hspace{7cm} B_0(\Mh^2,\MP,M_W)\\
\Delta T_{ 3,G^+G^-} &=& \frac{-iM_Z}{(8\cw)}\CZb\Sab
                         (1+\SQZb-\frac{\sw^2}{\cw^2}\CQZb)  \nonumber \\
                     & & \hspace{7cm} B_0(\Mh^2,M_W,M_W) \\
\Delta T_{ 4,\tilde{f}_R\tilde{f}_R} &=& \frac{i}{2} \bigg[ \frac{M_Z}{\cw}
         Q_f\sw^2\Sab - \frac{\mf^2}{M_W\Sb}\Ca \bigg]  \nonumber \\
         & & \quad \bigg[ Q_f \twz\CZb - \frac{\mf^2}{M_W^2}\TQb \bigg]
             N_f B_0(\Mh^2,\mfr,\mfr) \\
\Delta T_{ 4,\tilde{f'}_R\tilde{f'}_R} &=& \frac{i}{2} \bigg[ \frac{M_Z}{\cw}
         Q_{f'}\sw^2\Sab + \frac{\mfsf^2}{M_W\Cb}\Sa \bigg] \nonumber \\
         & & \quad \bigg[ Q_{f'} \twz\CZb - \frac{\mfsf^2}{M_W^2}\TQb \bigg]
             N_{f'} B_0(\Mh^2,\mfsr,\mfsr) \\
\Delta T_{ 5,\tilde{f}_L\tilde{f}_L} &=& \frac{i}{2} \bigg[ \frac{M_Z}{\cw}
       (\frac{1}{2}-Q_f\sw^2)\Sab - \frac{\mf^2}{M_W\Sb}\Ca \bigg]  \nonumber \\
       & & \quad \bigg[ \frac{\frac{1}{2}-Q_f\sw^2}{\cw^2}\CZb
           - \frac{\mf^2}{M_W^2}\TQb \bigg] N_f B_0(\Mh^2,\mfl,\mfl) \\
\Delta T_{ 5,\tilde{f'}_L\tilde{f'}_L} &=& \frac{i}{2} \bigg[ \frac{M_Z}{\cw}
           (\frac{1}{2}-Q_{f'}\sw^2)\Sab + \frac{\mfsf^2}{M_W\Cb}\Sa \bigg]
           \nonumber \\
       & & \quad \bigg[ \frac{\frac{1}{2}-Q_{f'}\sw^2}{\cw^2}\CZb
           - \frac{\mfsf^2}{M_W^2}\TQb \bigg] N_{f'} B_0(\Mh^2,\mfsl,\mfsl) \\
\Delta T_{ 6,Z^0Z^0} &=& \frac{iM_Z}{2\cw^3}\Sba\Big[ B_0(\Mh^2,M_Z,M_Z)
                          - \frac{1}{2} \Big] \\
\Delta T_{ 7,W^+W^-} &=& iM_W\Sba\Big[ 2B_0(\Mh^2,M_W,M_W) -1 \Big] \\
\Delta T_{ 8,h^0A^0} &=& \frac{iM_Z}{(2\cw)^3}\CQZb\Sab \CZa
                         B_0(\MA^2,\mh,\MA)\\
\Delta T_{ 8,h^0G^0} &=& \frac{iM_Z}{(2\cw)^3}\SQZb\Cab \CZa
                         B_0(\MA^2,\mh,M_Z)\\
\Delta T_{ 8,H^0A^0} &=& \frac{-iM_Z}{(2\cw)^3}\CQZb\Cab \SZa
                         B_0(\MA^2,\mH,\MA)\\
\Delta T_{ 8,H^0G^0} &=& \frac{-iM_Z}{(2\cw)^3}\SQZb\Cab \SZa
                         B_0(\MA^2,\mH,M_Z)\\
\Delta T_{ 9,H^0G^0} &=& \frac{iM_W}{8}\Sba B_0(\MA^2,\MP,M_W)\\
\Delta T_{10,A^0h^0h^0} &=& \frac{3iM_Z^3}{(2\cw)^3}\CQZb\SDab\CZa
                            C_0(\mh,\MA,\mh) \\
\Delta T_{10,A^0h^0H^0} &=& \frac{-iM_Z^3}{(2\cw)^3}\CQZb\Sab\Cab \nonumber \\
                        & & \quad (2\SZa\Sab-\Cab\CZa) C_0(\mh,\MA,\mH) \\
\Delta T_{10,A^0H^0H^0} &=& \frac{-iM_Z^3}{(2\cw)^3}\CQZb\CQab \nonumber \\
                        & & (2\SZa\Cab-\Sab\CZa)  C_0(\mH,\MA,\mH) \\
\Delta T_{10,G^0h^0h^0} &=& \frac{3iM_Z^3}{(2\cw)^3}\CQZb\SDab\CZa
                            C_0(\mh,M_Z,\mh) \\
\Delta T_{10,G^0h^0H^0} &=& \frac{iM_Z^3}{(2\cw)^3}\SQZb\Sab\Cab \nonumber \\
                        & & \quad (2\SZa\Sab-\Cab\CZa)  C_0(\mh,M_Z,\mH) \\
\Delta T_{10,G^0H^0H^0} &=& \frac{-iM_Z^3}{(2\cw)^3}\SQZb\CQab \nonumber \\
                        & & (2\SZa\Cab-\Sab\CZa)  C_0(\mH,M_Z,\mH) \\
\Delta T_{11,G^+H^+H^+} &=& \frac{iM_W^2}{4}\KL M_W\Sba+\frac{M_Z}{2\cw} \KR
                            \CZb\Sab \nonumber \\
                        & & \hspace{6cm} C_0(\MP,M_W,\MP) \\
\Delta T_{12,H^+G^+G^+} &=& \frac{-iM_Z^3}{(2\cw)^3} \CZb\Sab
                            C_0(\MP,M_W,\MP) \\
\Delta T_{13,\tilde{f}_L\tilde{f}_R\tilde{f}_R} &=& -i\KKL \frac{M_Z}{\cw}Q_{f}
        \sw^2\Sab - \frac{\mf^2\Ca}{M_W\Sb} \KKR \frac{\mf^2}{4M_W^2}(\mu -
        A_{f}\CTb)^2 \nonumber \\
        & & \hspace{5cm} N_f C_0(\mfr^2,\mfl^2,\mfr^2) \\
\Delta T_{13,\tilde{f'}_L\tilde{f'}_R\tilde{f'}_R} &=& -i\KKL \frac{M_Z}{\cw}
        Q_{f'} \sw^2
        \Sab + \frac{\mfsf^2\Sa}{M_W\Cb} \KKR \frac{\mfsf^2}{4M_W^2}(\mu -
        A_{f'}\Tb)^2 \nonumber \\
        & & \hspace{5cm} N_{f'} C_0(\mfsr^2,\mfsl^2,\mfsr^2) \\
\Delta T_{14,\tilde{f}_R\tilde{f}_L\tilde{f}_L} &=& -i\KKL \frac{M_Z}{\cw}
        \KL \frac{1}{2}-Q_{f} \sw^2 \KR
        \Sab - \frac{\mf^2\Ca}{M_W\Sb} \KKR \frac{\mf^2}{4M_W^2}(\mu -
        A_{f}\CTb)^2 \nonumber \\
        & & \hspace{5cm} N_f C_0(\mfsl^2,\mfsr^2,\mfsl^2) \\
\Delta T_{14,\tilde{f'}_R\tilde{f'}_L\tilde{f'}_L} &=& -i\KKL \frac{M_Z}{\cw}
        \KL \frac{1}{2}-Q_{f'} \sw^2 \KR
        \Sab - \frac{\mfsf^2\Ca}{M_W\Sb} \KKR \frac{\mfsf^2}{4M_W^2}(\mu -
        A_{f'}\Tb)^2 \nonumber \\
        & & \hspace{5cm} N_{f'} C_0(\mfsl^2,\mfsr^2,\mfsl^2) \\
\Delta T_{15,fff} &=& i\frac{\mf^4\Ca\CTQb}{(2M_W)^3\Sb}\Big\{ C_0(\mf,\mf,\mf)
                      [2\MA^2-\Mh^2] \nonumber \\
                  & & \hspace{3cm} -2B_0(\Mh^2,\mf,\mf)+1 \Big\} N_f \\
\Delta T_{15,f'f'f'} &=& i\frac{\mfsf^4\Sa\TQb}{(2M_W)^3\Cb}\Big\{
                         C_0(\mfsf,\mfsf,\mfsf) [2\MA^2-\Mh^2] \nonumber \\
                     & & \hspace{3cm} -2B_0(\Mh^2,\mfsf,\mfsf) \Big\} N_{f'}
\EEA

\vorn In the following, vertex corrections $U, V$ and $N$ denote the 
diagonalising matrices given in appendix A. The convention for $A, A_+, A_-$
etc.~are shown below ($\omega_\pm = \frac{1\pm\gamma_5}{2}$):

\epsfclipon
\begin{center}
\unitlength1mm
\begin{picture}(80,30)
%\put(0,0){\framebox(80,50)[t]{}}
\put(000,011){\epsffile[188 534 302 605]{top1a.eps}}
\put(12,10){\vector(0,1){10}}
\put(12,08){\makebox(0,0)[t]{$A(A_+\OP+A_-\OM)$}}
\put(40,28){\vector(-1,0){11}}
\put(41,28){\makebox(0,0)[l]{$B(B_+\OP+B_-\OM)$}}
\put(40,15){\vector(-1,0){11}}
\put(41,15){\makebox(0,0)[l]{$C(C_+\OP+C_-\OM)$}}
\end{picture}
\end{center}
\epsfclipoff

\BEA
\Delta T_{16,\tilde{\chi}^+_k\tilde{\chi}^+_i\tilde{\chi}^+_j}
&=& -ABC\Bigg\{ C_0(\mcpi,\mcpk,\mcpj) \\ \nonumber
 & & \qquad \Bigg[ \BL B_+A_+C_+ + B_-A_-C_-\BR \mcpi\mcpj\mcpk \\ \nonumber
 & & \qquad -\BL B_+A_-C_+ + B_-A_+C_-\BR \mcpk \frac{\Mh^2-\mcpi^2-\mcpj^2}
      {2} \\ \nonumber
 & & \qquad -\BL B_+A_+C_- + B_-A_-C_+\BR \mcpi \frac{\MA^2-\mcpk^2-\mcpj^2}
      {2} \\ \nonumber
 & & \qquad -\BL B_+A_-C_- + B_-A_-C_+\BR \mcpj \frac{\MA^2-\mcpi^2-\mcpk^2}
      {2} \Bigg] \\ \nonumber
 &+& B_0(\Mh^2,\mcpi,\mcpj) \KL \BL B_+A_+C_- + B_-A_-C_+\BR \frac{\mcpi}{2} +
                          \BL B_+A_-C_- + B_-A_+C_+\BR \frac{\mcpj}{2} \KR
                          \\ \nonumber
 &+& B_0(\MA^2,\mcpk,\mcpj) \KL \BL B_+A_-C_+ + B_-A_+C_-\BR \frac{\mcpk}{2} +
                            \BL B_+A_-C_- + B_-A_+C_+\BR \frac{\mcpj}{2} \KR
                          \\ \nonumber
 &+& B_0(\MA^2,\mcpi,\mcpj) \KL \BL B_+A_+C_- + B_-A_-C_+\BR \frac{\mcpi}{2} +
                            \BL B_+A_-C_+ + B_-A_+C_-\BR \frac{\mcpj}{2} \KR
                          \Bigg \}
\EEA

\vorn The following couplings have to be inserted:
\BEA
A &=& i, \quad A_+=Q_{ij}\Sa-S_{ij}\Ca, \quad A_-=Q_{ji}^*\Sa-S_{ji}^*\Ca
      \nonumber \\
B &=& -g, \quad B_+=Q_{ki}\Sb-S_{ki}\Cb, \quad B_-=Q_{ik}^*\Sb-S_{ik}^*\Cb
      \nonumber \\
C &=& -g, \quad C_+=Q_{jk}\Sb-S_{jk}\Cb, \quad C_-=Q_{kj}^*\Sb-S_{kj}^*\Cb
      \nonumber \\
Q_{ij} &=& \sqrt{\frac{1}{2}}V_{i1}U_{j2} \nonumber \\
S_{ij} &=& \sqrt{\frac{1}{2}}V_{i2}U_{j1} \nonumber
\EEA

\BEA
\Delta T_{17,\tilde{\chi}^0_k\tilde{\chi}^0_i\tilde{\chi}^0_j}
&=& -ABC\Bigg\{ C_0(\mcni,\mcnk,\mcnj) \\ \nonumber
 & & \qquad \Bigg[ \BL B_+A_+C_+ + B_-A_-C_-\BR \mcni\mcnj\mcnk \\ \nonumber
 & & \qquad -\BL B_+A_-C_+ + B_-A_+C_-\BR \mcnk \frac{\Mh^2-\mcni^2-\mcnj^2}
      {2} \\ \nonumber
 & & \qquad -\BL B_+A_+C_- + B_-A_-C_+\BR \mcni \frac{\MA^2-\mcnk^2-\mcnj^2}
      {2} \\ \nonumber
 & & \qquad -\BL B_+A_-C_- + B_-A_-C_+\BR \mcnj \frac{\MA^2-\mcni^2-\mcnk^2}
      {2} \Bigg] \\ \nonumber
 &+& B_0(\Mh^2,\mcni,\mcnj) \KL \BL B_+A_+C_- + B_-A_-C_+\BR \frac{\mcni}{2} +
                          \BL B_+A_-C_- + B_-A_+C_+\BR \frac{\mcnj}{2} \KR
                          \\ \nonumber
 &+& B_0(\MA^2,\mcnk,\mcnj) \KL \BL B_+A_-C_+ + B_-A_+C_-\BR \frac{\mcnk}{2} +
                            \BL B_+A_-C_- + B_-A_+C_+\BR \frac{\mcnj}{2} \KR
                          \\ \nonumber
 &+& B_0(\MA^2,\mcni,\mcnj) \KL \BL B_+A_+C_- + B_-A_-C_+\BR \frac{\mcni}{2} +
                            \BL B_+A_-C_+ + B_-A_+C_-\BR \frac{\mcnj}{2} \KR
                          \Bigg\}
\EEA

\vorn The following couplings have to be inserted:
\BEA
A &=& i, \quad A_+=Q_{ij}''\Sa-S_{ij}''\Ca, \quad
      A_-=Q_{ji}^{\prime\prime *}\Sa-S_{ji}^{\prime\prime *}\Ca \nonumber \\
B &=& -g, \quad B_+=Q_{ki}''\Sb-S_{ki}''\Cb, \quad
      B_-=Q_{ik}^{\prime\prime *}\Sb-S_{ik}^{\prime\prime *}\Cb \nonumber \\
C &=& -g, \quad C_+=Q_{jk}''\Sb-S_{jk}''\Cb, \quad
      C_-=Q_{kj}^{\prime\prime *}\Sb-S_{kj}^{\prime\prime *}\Cb \nonumber \\
Q_{ij}'' &=& \frac{1}{2}[N_{i3}(N_{j2}-N_{j1}\tan\theta_W)+
             N_{j3}(N_{i2}-N_{i1}\tan\theta_W)] \nonumber \\
S_{ij}'' &=& \frac{1}{2}[N_{i4}(N_{j2}-N_{j1}\tan\theta_W)+
             N_{j4}(N_{i2}-N_{i1}\tan\theta_W)] \nonumber
\EEA

\BEA
\Delta T_{18,h^0Z^0A^0} &=& \frac{2iM_Z}{(2\cw)^3}\CQba\CZb\Sab \\
        & & \Big\{ \quad C_0(M_Z,\mh,\MA)[\mh^2-M_Z^2+\Mh^2] \nonumber \\
        & & \qquad +B_0(\MA^2,\mh,M_Z)-B_0(\MA^2,\mh,\MA)+B_0(\MH^2,M_Z,\MA)
            \Big\} \nonumber \\
\Delta T_{18,H^0Z^0A^0} &=& \frac{2iM_Z}{(2\cw)^3}\Cba\CZb\Sba\Cab \\
        & & \Big\{ \quad C_0(M_Z,\mH,\MA)[-M_Z^2+\Mh^2+\mH^2] \nonumber \\
        & & \qquad +B_0(\MA^2,\mH,M_Z)-B_0(\MA^2,\mH,\MA)+B_0(\Mh^2,M_Z,\MA)
            \Big\} \nonumber \\
\Delta T_{18,h^0Z^0G^0} &=& \frac{2iM_Z}{(2\cw)^3}\Cba\SZb\Sba\Sab \\
        & & \Big\{ \quad C_0(M_Z,\mh,M_Z)[-\MA^2+2\Mh^2]
            +B_0(\Mh^2,M_Z,M_Z) \Big\} \nonumber \\
\Delta T_{18,H^0Z^0G^0} &=& \frac{2iM_Z}{(2\cw)^3}\SZb\SQba\Cab \\
        & & \Big\{ \quad C_0(M_Z,\mH,M_Z)[-\MA^2+\Mh^2+\MH^2]
            +B_0(\Mh^2,M_Z,M_Z) \Big\} \nonumber \\
\Delta T_{19,H^+W^+G^+} &=& \frac{-2iM_W}{8}\Sba \\
        & & \Big\{ \quad C_0(M_W,\MP,M_W)[\MA^2-\Mh^2-\MP^2]
            -B_0(\Mh^2,M_W,M_W) \Big\} \nonumber \\
\Delta T_{20,Z^0H^0h^0} &=& \frac{-iM_Z}{(2\cw)^3}\Cba\Sba   
            (2\SZa\Sab-\Cab\CZa) \nonumber \\
        & & \Big\{ \quad C_0(\mH,M_Z,\mh)[M_Z^2-2\MA^2-\MH^2+2\Mh^2-\mh^2] \\
        & & \qquad +B_0(\Mh^2,\mH,\mh)-B_0(\MA^2,M_Z,\mH)-B_0(\MA^2,M_Z,\mh)
            \Big\} \nonumber \\
\Delta T_{20,Z^0h^0h^0} &=& \frac{3iM_Z}{(2\cw)^3}\CQba\Sab\CZa \\
        & & \Big\{ \quad C_0(\mh,M_Z,\mh)[M_Z^2-2\MA^2+2\Mh^2-2\mh^2] 
            \nonumber \\
        & & \qquad +B_0(\Mh^2,\mh,\mh)-2B_0(\MA^2,M_Z,\mh) \Big\} \nonumber \\
\Delta T_{20,Z^0H^0H^0} &=& \frac{-iM_Z}{(2\cw)^3}\SQba
            (2\SZa\Sab-\Cab\CZa) \nonumber \\
        & & \Big\{ \quad C_0(\mH,M_Z,\mH)[M_Z^2-2\MA^2-2\mH^2+2\Mh^2] \\
        & & \qquad +B_0(\Mh^2,\mH,\mH)-2B_0(\MA^2,M_Z,\mH) \Big\} \nonumber \\
\Delta T_{20,Z^0h^0H^0} &=& \frac{-iM_Z}{(2\cw)^3}\Cba\Sba   
            (2\SZa\Sab-\Cab\CZa) \nonumber \\
        & & \Big\{ \quad C_0(\mh,M_Z,\mH)[M_Z^2-2\MA^2-\mH^2+2\Mh^2-\mh^2] \\
        & & \qquad +B_0(\Mh^2,\mh,\mH)-B_0(\MA^2,M_Z,\mH)-B_0(\MA^2,M_Z,\mh)
            \Big\} \nonumber \\
\Delta T_{21,W^+H^+H^+} &=& \frac{i}{2}\KL M_W\Sba+\frac{M_Z}{2\cw}\CZb\Sab
            \KR \\
        & & \Big\{ \quad C_0(\MP,M_W,\MP)[M_W^2-2\MA^2-2\MP^2+2\Mh^2]   
            \nonumber \\
        & & \qquad +B_0(\Mh^2,\MP,\MP)-2B_0(\MA^2,M_W,\MP) \Big\} \nonumber \\
\Delta T_{22,H^0Z^0Z^0} &=& \frac{2iM_Z}{(2\cw)^3}\SDba \\
        & & \Big\{ \quad C_0(M_Z,\mH,M_Z)\KKL -\MA^2-\MH^2
            +\frac{1}{2}\Mh^2\KKR \nonumber \\
        & & \qquad -2B_0(\Mh^2,M_Z,M_Z)+B_0(\MA^2,\mH,M_Z) \Big\} \nonumber \\
\Delta T_{22,h^0Z^0Z^0} &=& \frac{2iM_Z}{(2\cw)^3}\Sba\CQba \\
        & & \Big\{ \quad C_0(M_Z,\mH,M_Z)\KKL -\MA^2-\mh^2
            +\frac{1}{2}\Mh^2\KKR \nonumber \\
        & & \qquad -2B_0(\Mh^2,M_Z,M_Z)+B_0(\MA^2,\mh,M_Z) \Big\} \nonumber \\
\Delta T_{23,H^+W^+W^+} &=& \frac{iM_W}{2}\Sba \\
        & & \Big\{ \quad C_0(M_W,\MP,M_W)\KKL -\MA^2-\MP^2+\frac{1}{2}\Mh^2\KKR
            \nonumber \\
        & & \qquad -2B_0(\Mh^2,M_W,M_W)+B_0(\MA^2,\MP,M_W) \Big\} \nonumber \\
\Delta T_{24,h^0A^0A^0} &=& \frac{iM_Z^3}{(2\cw)^3}\CDZb\SDab
                            C_0(\MA,\mh,\MA) \\
\Delta T_{24,h^0A^0G^0} &=& \frac{iM_Z^3}{(2\cw)^3}\SQZb\SDab\CZb
                            C_0(\MA,\mh,M_Z) \\
\Delta T_{24,h^0G^0A^0} &=& \frac{iM_Z^3}{(2\cw)^3}\SQZb\SDab\CZb
                            C_0(M_Z,\mh,\MA) \\
\Delta T_{24,h^0G^0G^0} &=& \frac{-iM_Z^3}{(2\cw)^3}\SQZb\SDab\CZb
                            C_0(M_Z,\mh,M_Z) \\
\Delta T_{24,H^0A^0A^0} &=& \frac{iM_Z^3}{(2\cw)^3}\CDZb\CQab\Sab
                            C_0(\MA,\mH,\MA) \\
\Delta T_{24,H^0A^0G^0} &=& \frac{iM_Z^3}{(2\cw)^3}\SQZb\CQab\Sab\Cab 
                            \nonumber \\
                        & & \hspace{5cm} C_0(\MA,\mH,M_Z) \\
\Delta T_{24,H^0G^0A^0} &=& \frac{iM_Z^3}{(2\cw)^3}\SQZb\CQab\Sab\Cab
                            \nonumber \\
                        & & \hspace{5cm} C_0(M_Z,\mH,\MA) \\
\Delta T_{24,H^0G^0G^0} &=& \frac{-iM_Z^3}{(2\cw)^3}\SQZb\CQab\Sab\CZb
                            C_0(M_Z,\mH,M_Z) 
\EEA

\end{appendix}

%\bibitem{} , Nucl. Phys. {\bf B  22} (1961) 579
%\bibitem{} , Phys. Lett. {\bf  12} (1964) 132
%\bibitem{} , Phys. Lett. {\bf B 12} (1964) 132
%\bibitem{} , Phys. Rep. {\bf  20} (1970) 1285
%\bibitem{} , Phys. Rev. {\bf  20} (1970) 1285
%\bibitem{} , Phys. Rev. {\bf D 20} (1970) 1285
%\bibitem{} , Phys. Rev. Lett. {\bf  19} (1967) 19
%\bibitem{} , Progr. Theor. Phys. {\bf 49} (1973) 652

\end{document}